\documentclass[12pt]{article}
\usepackage{authblk}
\usepackage[bookmarksnumbered, colorlinks, plainpages]{hyperref}
\usepackage{braket}
\usepackage{amsmath, amsthm, amscd, amsfonts, amssymb, graphicx, color, booktabs}
\usepackage{cancel}
\usepackage{xspace}
\usepackage{subfigure}
\usepackage{cite}
\newcommand{\re}{{\rm Re}}
\newcommand{\im}{{\rm Im}}

\makeatletter
\ifcase \@ptsize \relax
  \newcommand{\miniscule}{\@setfontsize\miniscule{4}{5}}
\or
  \newcommand{\miniscule}{\@setfontsize\miniscule{5}{6}}
\or
  \newcommand{\miniscule}{\@setfontsize\miniscule{5}{6}}
\fi
\makeatother

\DeclareRobustCommand{\optbar}[1]{\shortstack{{\miniscule (\rule[.5ex]{1.25em}{.18mm})}
  \\ [-.7ex] $#1$}}

\numberwithin{equation}{section}
\definecolor{email}{rgb}{0.00,0.00,0.84}
\begin{document}
\setcounter{page}{1}

\title{WG5 summary: Direct CP violation (DCPV) including $\phi_{3}/\gamma$ from $B\to DK$, DCPV effects, branching fractions and polarisation in charmless $B_{(s)}$ decays}

\author[1]{Melissa Cruz Torres}
\author[2]{Tobias Huber}
\author[3]{Minakshi Nayak}
\author[2]{Gilberto Tetlalmatzi-Xolocotzi}

\affil[1]{Universidad Nacional Autónoma de Honduras: Tegucigalpa, Francisco Morazán, Honduras}
\affil[2]{Theoretische Physik 1, Center for Particle Physics Siegen (CPPS), Universit\"aat Siegen, Walter-Flex-Strasse 3, D-57068 Siegen, Germany.}
\affil[3]{Center for High Energy Physics, Indian Institute of Science, CV Raman Rd, Bengaluru, 560012, India}

\maketitle             

\begin{abstract}
\noindent In this contribution a summary of the activities of Working Group 5 (WG5) presented during the 12th International Workshop on the CKM Unitarity Triangle (CKM2023) is reported. This includes new results on $\phi_{3}/\gamma$ measurements using $B\to DK$ decays, search for $CP$ violation using charmless $B$ decays and $b$-Baryon decays, measurement of branching ratios in hadronic $B$ to charm decays, and theory of three-body nonleptonic $B$ decays.
\end{abstract} \maketitle

\vspace*{40pt}

{\footnotesize{Preprint: SI-HEP-2024-12, P3H-24-030}}\normalsize

\newpage

\section{Introduction}

\noindent

Non-leptonic $B$ meson decays are of capital importance in testing our understanding of the Standard Model (SM) and beyond. To begin with, there is currently a plethora of experimental data from multiple sources including B factories, the Tevatron and the LHC. Moreover, they are ideally suited to probe for CP violation. Indeed, in the SM, nonleptonic $B$ meson processes give us sensitivity to the CKM weak phase at tree level in channels such as $B^-\rightarrow \bar{D}^0 (D^0) K^-$. This is particularly advantageous over for instance, semileptonic processes such as $B\rightarrow K^{(*)} \ell \ell$ \cite{Bobeth:2008ij,Bobeth:2011gi,Descotes-Genon:2015hea,Fleischer:2022klb} where the $b\rightarrow s$ asymmetries are doubly Cabibbo suppressed, or even with respect of leptonic decays $B\rightarrow \ell^+ \ell^-$ where the sensitivity towards CP violation arises from neutral $B_{d,s}$ meson mixing and is therefore also expected to be tiny in the SM \cite{DeBruyn:2012wk,Fleischer:2017yox,Buras:2013uqa}. 

One of the issues in nonleptonic processes are the large uncertainties on the theory side which affect the decay amplitudes beyond leading order in the $\Lambda_{\mathrm{QCD}}/m_b$ expansion and which appear for instance in annihilation topologies. Nevertheless, different approaches can be pursued in this direction, including the identification of channels where annihilation is absent \cite{Bordone:2020gao}, using data to obtain bounds for the power corrections \cite{Huber:2021cgk}  and constructing observables with low sensitivity to them \cite{Biswas:2024bhn, Biswas:2023pyw,Alguero:2020xca}.

In these proceedings, we summarise the recent studies of direct $CP$ violation in hadronic $B$ decays and measurement of branching ratios in hadronic $B$ to charm decays as presented at the CKM conference 2023. The results on direct $CP$ violation are divided into three categories: measurements of $CP$ violating parameter $\phi_3/\gamma$ from the $B \to DK$ decays; studies of direct $CP$ violation in charmless decays; searches for $CP$ violation in b-baryons.  We also discuss the current status of $B$ meson three body decays within the framework of QCD factorization for Heavy-Light-Light and in the situation where there are three light mesons in the final states.

\section{$\phi_3$/$\gamma$ measurements }
The angle $\phi_3$, also know as $\gamma$, which is expressed in terms of CKM matrix elements as $\arg(-V_{ud}V_{ub}^*/V_{cd}V_{cb}^*)$ is the only CKM angle accessible through tree level decays via interference between favoured $b\to c$ and suppressed $b\to u$ decay amplitudes. The ideal tree decays to measure $\gamma$ is $B \to DK$ which has clean background, large branching fraction. Since no $B$-mixing nor penguin amplitudes are involved, and the tree level nature of the amplitudes involved in $B \to DK$ decays allows the theoritical clean extraction of $\gamma$. Here $D$ represents a general superposition of $D^0$ and $\bar{D}^0$ states.

Extraction of $\gamma$ involves measurement of $B^- \to \bar{D^{0}}K^-$ and  $B^- \to D^{0}K^-$  amplitudes, where the former one is both CKM and colour suppressed. The ratio of the suppressed to favoured amplitudes is written as: 
\begin{eqnarray}
\frac{\mathcal{A}(B^-\to \bar{D^{0}}K^-)}{\mathcal{A}(B^-\to D^{0}K^-)} & = & r_{B}e^{i(\delta_B - \gamma)},\
\label{eq:ratio1}
\end{eqnarray}
where $r_B \approx 0.1$ is the ratio of magnitudes and $\delta_B$ is the strong-phase difference. Since the hadronic parameters $r_B$ and $\delta_B$ can be determined from data together with $\phi_3$ makes these measurements are essentially free of theoretical uncertainties~\cite{Brod:2013sga}.

Several methods using different $D$ final states accessible to both $D^0$ and $\bar{D^0}$ are utilized to extract $\gamma$: GLW method ~\cite{glw} which uses $D$ decay to $CP$-eigenstate, such as $K^+K^-, \pi^+\pi^-, K^0_{S}\pi^0$; ADS method~\cite{ads} which uses flavor state such as $D \to K^\pm \pi^\mp$;  BPGGSZ method~\cite{Giri:2003ty, Bondar:2005ki, Grossman:2002aq} which uses self-conjugate multibody state of $D$ such as $K^0_S\pi^-\pi^+$. Other possibilities include the decays of neutral $B$ mesons~\cite{Kayser:1999bu, Gronau:2004gt}, multibody $B$ decays~\cite{Aleksan:2002mh, Gershon:2008pe, Gershon:2009qr, Gershon:2009qc} and $D^*$ or $D^{**}$ decays~\cite{Bondar:2004bi, Sinha:2004ct}. The best sensitivity is achieved by combining the various methods and decay modes as each of them has a different sensitivity to $\gamma$ depending on the $CP$-conserving parameters entering the decay amplitudes. The current world-average value of $\gamma$ from the direct measurements using tree-level decays is $(65.9^{+3.3}_{-3.5})^{\circ}$~\cite{HFLAV:2022esi}, whereas indirect determination of $\gamma$ from measurement of $\alpha$ and $\beta$ is $(66.29^{+0.72}_{-1.86})^{\circ}$~\cite{phi3_indirect}. Hence, improvement in the direct measurement is necessary to constrain any possible new physics contributions. 

It can be shown that in case of GLW method, the decay rates of $B^\pm$ 
\begin{equation}
\Gamma(B^\pm \to D_{CP}K^\pm)  =  |1 + r_{B}e^{-i(\delta_B \pm \gamma)}|^2  =  1+ {r^2_B}\pm 2 \kappa r_B \cos{(\delta_B \pm \gamma)} \, ,
\label{eq:ratio2}
\end{equation}
%
%
%
where the factor $\kappa = (2F_+ -1)$ =1 for two-body $D$ decays, but for multi-body $D$ decays, one needs charm factory input to determine the $CP$ even content $F_+$.

In case of the ADS method which includes non-$CP$ eigenstates of $D$ decays, a factor relating $D^0$ and $\bar D^0$ to a common final state, $r_{D}e^{i\delta_D}$ is included. In the charm meson decay only a $CP$-conserving phase is needed as $CP$-violating effects are neglected. Hence, the decay rates of $B$ is expressed as
\begin{equation}
\Gamma(B \to D_{\rm flav}K)  =  |r_{D}e^{i\delta_D} + r_{B}e^{-i(\delta_B \pm \gamma)}|^2  =  r^2_{D}+ {r^2_B}\pm 2 R_{f} r_D r_B \cos{(\delta_B - \delta_D \pm \gamma)} \, ,
\label{eq:ratio5}
\end{equation}
where the coherence factor $R_f =1 $ for two-body $D$ decays. Multi-body decays need input from charm factory to determine the coherence factor $R_f$. 
\subsection{Inputs for $\gamma$ measurements from BESIII}
Improving the precision of $\gamma$ measurement at LHCb/Belle II requires the external information on the hadronic parameters of $D$ meson decays that is produced from the $B$ decay. BESIII data from entangled $\psi(3770) \to D \bar D$ decays, provide a unique access to these strong parameters. The updated measurement of the parameter $\delta^{K\pi}_{D}$ for $D \to K\pi$ using the current dataset of $2.9~\rm fb^{-1}$ at BESIII is reported~\cite{BESIII:2022qkh}. This is the most precise measurement of $\delta^{K\pi}_{D}$ in quantum correlated $D\bar D$ decays. Measurement of $R_{K3\pi}$, $\delta_{D}^{K3\pi}$ for $D \to K^{-}\pi^{+}\pi^{-}\pi^{+}$ and $R_{K\pi\pi^0}$, $\delta_{D}^{K\pi\pi^0}$ for $D \to K^{-}\pi^{+}\pi^0$ from BESIII are also presented~\cite{BESIII:2021eud}. BESIII also reported measurements of $CP$-even fraction of $D \to \pi^+ \pi^- \pi^+ \pi^-$, $D \to K^+ K^- \pi^+ \pi^-$, $D \to K_{S} \pi^- \pi^+ \pi^0$ decays~\cite{BESIII:2022wqs, BESIII:2022ebh, BESIII:2023xgh}. These measurements are used as external inputs for $\gamma$ measurements at LHCb discussed below.
\subsection{New results of $\gamma$ measurement in ADS and GLW-like decays at LHCb}
\begin{figure}[htb]
\includegraphics[width=0.5
\linewidth]{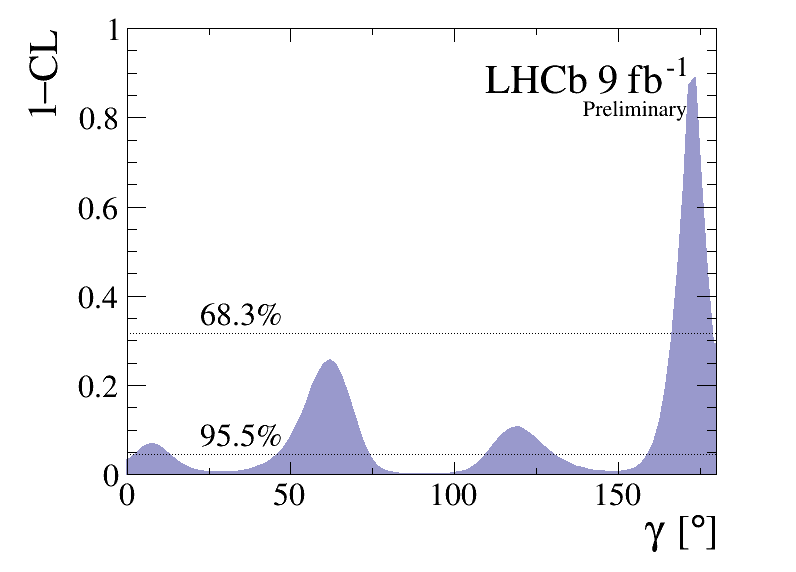}
\includegraphics[width=0.5
\linewidth]{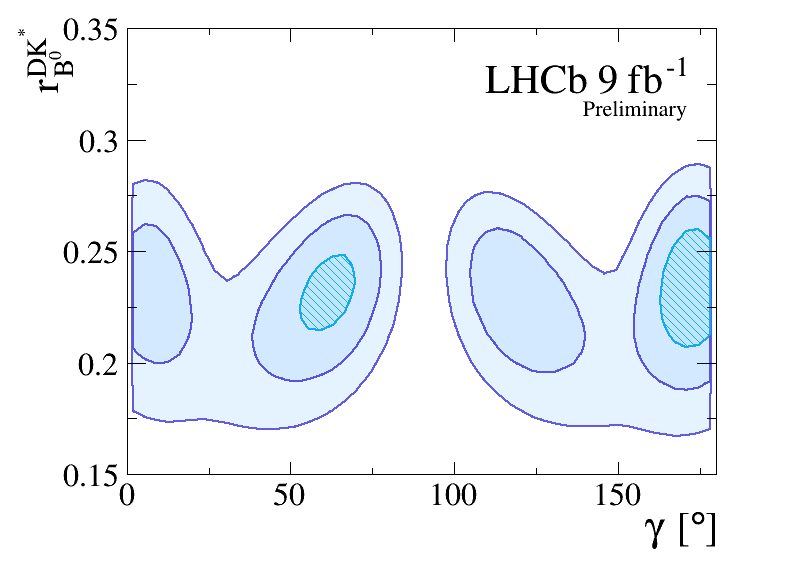}
\caption{Confidence level contours from the fitted results of the $B^0$-related $CP$-violating observables projected to the $\gamma$ axis (left), and to the $\gamma-r^{DK^*}_{B^0}$ plane (right).}
\label{fig:gamma-rb} 
\end{figure}

The current average value of $\gamma$ from LHCb Collaboration is $(63.8^{+3.5}_{-3.7})^o$~\cite{LHCb:conf2022-003} which is the most precise
determination from a single experiment.
In this conference, LHCb reported new results using $B^0 \to DK^{*}(892)^0$~\cite{LHCb:conf2023-003}. Despite the smaller branching fraction of $B^0 \to DK^{*}(892)^0$ decays, a competitive precision on $\gamma$ can be achieved due to the larger interference effects in these decays compared to the $B^+$ decays~\cite{Gershon:2008pe}. The analysis reconstructs $K^{*}(892)^0 \to K^+\pi^-$ and $D \to K\pi(\pi\pi), \pi\pi(\pi\pi), KK$ final states using proton-proton data collected by the LHCb experiment corresponding to an integrated luminosity of $9~\rm fb^{-1}$. The $CP$ violating observables such as ratios of branching fractions and charge asymmetries measured for the GLW modes $D \to KK, \pi\pi\pi\pi$ and for the ADS modes $D \to \pi K, \pi K \pi\pi$ are

\begin{eqnarray}
A^{KK}_{CP}  &=&  -0.047 \pm 0.063 \pm 0.015,\\
R^{KK}_{CP}  &=&  0.817 \pm 0.057 \pm 0.017,\\
R^{4\pi}_{CP}  &=&  0.882 \pm 0.086 \pm 0.033,\\
A^{4\pi}_{CP}  &=&  0.014 \pm 0.087 \pm 0.016,\\
R^{+}_{\pi K}  &=&  0.069 \pm 0.013 \pm 0.005,\\
R^{-}_{\pi K}  &=&  0.093 \pm 0.013 \pm 0.005,\\
R^{+}_{\pi K\pi\pi}  &=&  0.060 \pm 0.014 \pm 0.005,\\
R^{-}_{\pi K\pi\pi}  &=&  0.038 \pm 0.014 \pm 0.005,
\label{eq:ratio6}
\end{eqnarray}
where the uncertainties are statistical and systematic, respectively. These observables along with some additional inputs from other charm experiments are used to set constraints on the parameter space of $\gamma$  and the hadronic parameters $r^{DK^*}_{B0}$ and $\delta^{DK^*}_{B0}$. Out of four solutions of $\gamma$, the preferred solution is $\gamma = (172 \pm 6)^o$, and the second preferred solution of $\gamma = (62 \pm 8)^{o}$ is most consistent with world-average of direct measurements. The result of $B^0 \to D[K^{0}_{S} h^+ h^-]K^{*0}$~\cite{LHCb:conf2023-009}
break the degeneracy of these two solutions, indicating a strong preference for the second solution which is consistent with the world-average.
$\gamma$ using fully and partially reconstructed $B^\pm \to D^{*}K^{\pm}$ with $D \to K^{0}_{S} h^+ h^-$ was also measured by LHCb. Using fully reconstructed $B^\pm$, the measured value of $\gamma$ is $(69^{+13}_{-14})^{o}$. The value from partially reconstructed $B^\pm$ is $\gamma = (92^{+21}_{-17})^{o}$, where the uncertainty is statistically dominated, and the result is consistent with expectations. Confidence level contours from the fitted results of the $B^0$-related $CP$-violating observables projected to the $\gamma$ axis, and $\gamma - r^{DK^{*0}}_{B^0}$ plane are shown in Fig.~\ref{fig:gamma-rb}.

Current LHCb average value of $\gamma = (63.8^{+3.5}_{-3.7})^o$, has a contribution from systematic uncertainties approximately $1.4^o$~\cite{LHCb:conf2022-003}. By the end of Run4, sub-degree level precision on $\gamma$ is expected ~\cite{Cerri:2018ypt} through direct measurements. But the current level of
systematic uncertainty will limit the precision measurements in future. So the total systematic uncertainty must be very well examined and needs to be reduced in order to achieve the sub-degree precision. Systematic uncertainties from different $B$ decays, and from LHCb vs $B$ factories  are largely uncorrelated, except the strong correlation due to shared inputs of $D$ strong phase inputs. BESIII will have seven times more data to provide precision measurements of strong phase inputs  in order to achieve this sub-degree precision of $\gamma$.

\subsection{New results of $\phi_3$ at Belle/Belle II }
Belle and Belle II collaborations using $711~\rm fb^{-1}$ of Belle data and $128~\rm fb^{-1}$ of Belle II data reported the CKM angle $\phi_3$ using $B \to D[K^{0}_{S} h^+ h^-]h^{-}$~\cite{Belle:2021efh}. The measured value of $\phi_3$ is $(78.4 \pm 11.4 \pm 0.5 \pm 1.0)^{o}$ compared to the previous measurement of $\phi_3 = (77.3^{+15.1}_{-14.9} \pm 4.1 \pm 4.3)^{o}$~\cite{Belle:2012ftx} by Belle using $711~\rm fb^{-1}$ of data. Here the first uncertainty is statistical, the second is the experimental systematic uncertainty and the third is from the uncertainties on external measurements of the $D$-decay strong-phase parameters. The improved statistical precision is due to improved $K_{S}^0$ selection, analysis strategy and background suppression compared to the previous one. The improved experimental systematic is primarily due to the improved background suppression and the use of the $B^+ \to D\pi^+$ sample to determine the acceptance. And the uncertainty related to $D$-decay strong-phase inputs has also decreased because of the new measurements reported by the BESIII collaboration~\cite{BESIII:2020khq, BESIII:2020hpo}.

Recently, they also reported GLW study for the decays $B \to D[KK, K^{0}_{S} \pi^0]K$ using  $711~\rm fb^{-1}$ of Belle data and $189~\rm fb^{-1}$ of Belle II data ~\cite{Belle-II:2023mwi}. The $CP$ asymmetries and the ratios of branching fractions measured for the above modes are 
\begin{eqnarray}
A_{CP+}  &=&  (+12.5 \pm 5.8 \pm 1.4)\%,\\
A_{CP-}  &=&  (-16.7 \pm 5.7 \pm 0.6)\%,\\
R_{CP+}  &=&  1.164 \pm 0.081 \pm 0.036,\\
R_{CP-}  &=&  1.151 \pm 0.074 \pm 0.019.
\label{eq:belle_glw1}
\end{eqnarray}
Belle and Belle II Collaborations also made GLS study for $B^{\pm} \to D[K^{0}_{S}K^{\pm}\pi^{\mp}]K^{\pm}$ and  $B^{\pm} \to D[K^{0}_{S}K^{\pm}\pi^{\mp}]\pi^{\pm}$ using  $711~\rm fb^{-1}$ of Belle data and $362~\rm fb^{-1}$ of Belle II data ~\cite{Belle:2023lha}. The decays are categorised as same-sign (SS) or opposite- sign (OS) according to the charge of the $K^\pm$ produced by the $D$ meson relative to the charge of the $B^\pm$ meson. Asymmetries and branching-fraction ratios obtained are 
\begin{eqnarray}
A^{DK}_{SS}  &=&  -0.089 \pm 0.091 \pm 0.011,\\
A^{DK}_{OS}  &=&  0.109 \pm 0.133 \pm 0.013,\\
A^{D\pi}_{SS}  &=&  0.018 \pm 0.026 \pm 0.009,\\
A^{D\pi}_{OS}  &=&  -0.028 \pm 0.031 \pm 0.009,\\
R^{DK/D\pi}_{SS}  &=&  0.122 \pm 0.012 \pm 0.004,\\
R^{DK/D\pi}_{OS}  &=&  0.093 \pm 0.013 \pm 0.003,\\
R^{D\pi}_{SS/OS}  &=&  1.428 \pm 0.057 \pm 0.002.
\label{eq:belle_glw2}
\end{eqnarray}
 
\begin{figure}[htb]
\includegraphics[width=0.9
\linewidth]{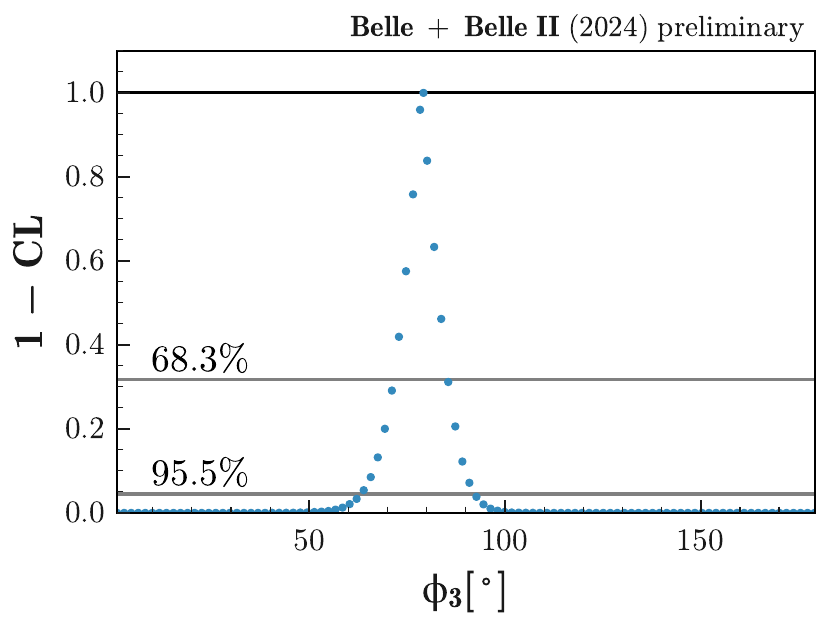}
\caption{One dimensional 1-CL scan showing the result of the $\phi_3$ combination.}
\label{fig:phi3-belleii} 
\end{figure}

In summary, both the collaborations recently reported a first combined determination of $\phi_3$ by combining existing Belle and Belle II measurements which includes inputs from four different methods (BPGGSZ, ADS, GLW, GLS) with 17 different $B^\pm \to D^{(*)}h^{\pm}$ final states and auxiliary $D$ decay informations from other experiments such as BESIII, CLEO, LHCb.  The first Belle + Belle II combination for $\phi_3$ is $(78.6^{+7.2} _{-7.3})^{o}$~\cite{Belle:2024eob}, consistent with world average value within $2\sigma$~\cite{HFLAV:2022esi}.
1-CL distribution as function of $\phi_3$ is shown in Fig.~\ref{fig:phi3-belleii}.

\section{Direct $CP$ Violation measurement in charmless b-hadrons}
\subsection{Theory of three-body non-leptonic $B$ decays}

Non-leptonic three-body decays of $B$ mesons can provide valuable information on $CP$ violation, the structure of QCD in processes involving heavy quarks, as well as potential New Physics effects in the quark sector.
On the theory side these processes have been investigated following different paths, including flavour symmetries \cite{Bhattacharya:2014eca,Bhattacharya:2023pef}, pQCD \cite{Wang:2014qya} as well as  model dependent approaches \cite{Bediaga:2017axw, Mannel:2020abt}. Here we focus on the theory developments  within the context of QCD factorization (QCDF) \cite{Beneke:1999br,Beneke:2000ry,Beneke:2001ev,Beneke:2003zv}. At leading power in $\Lambda_{\mathrm{QCD}}/m_b$ the formalism of QCDF allows us to decompose the corresponding amplitudes in terms of local form factors, (di)meson light-cone distribution amplitudes (LCDAs) and hard kernels which are process-dependent but perturbatively calculable. The study of three-body decays in QCDF is relatively new and has been put forward in~\cite{Beneketalk,Stewarttalk,Krankl:2015fha,Huber:2020pqb,Klein:2017xti}. Since for three-body final states the phase space is two-dimensional, the precise form of the factorization of the amplitudes depends on the region of phase space under consideration. In particular in the edges of the phase space, where one daughter-particle recoils against the other two and which therefore includes the resonant region where quasi-two-body decays such as $B\rightarrow \pi K^{\ast}$ and $B\rightarrow D \rho$ occur, the situation is similar to the one for two-body final states. Actually, if we consider the decay process $B\rightarrow M M_1 M_2$, the treatment of the region where the invariant mass of the $M_1 M_2$-pair is small is similar to the one given by two-body processes. The key differences stem from the form factors and the dimeson LCDAs.

\subsubsection{Heavy-light-light final states}

The simplest situation corresponds to the processes $B\rightarrow D M_1 M_2$, where the $D$ meson picks up the spectator quark of the initial $B$ meson and the combined system $(M_1 M_2)$, integrated by two light mesons $M_1$ and $M_2$, has a small invariant mass. More concretely,  we can develop a factorization formula for the decays 
\begin{eqnarray}
 \bar{B}^0 \rightarrow D^+ M^- \pi^0   
\end{eqnarray}
for $(M=K, \pi)$ in the phase space region where the $M^- \pi^0$ system recoils against the heavy $D$ meson. To perform the kinematic description we label the momenta of the $B^0$,  $D^+$,  $M^-$ and the $\pi^0$ mesons as $p$, $q$, $k_1$ and $k_2$ respectively. Then the phase space can be described in terms of two variables, the invariant mass of the system $(M\pi)$ which is given by $k^2=(k_1 + k_2)^2$ and the angle $\theta_{\pi}$ between the three momentum $\Vec{k}_2$ and $\Vec{p}$ in the $(M\pi)$ rest frame, where $\Vec{k}=0$ holds.

The decay amplitudes can be computed as 
\begin{eqnarray}
\mathcal{A}(\bar{B}^0 \to D^{+}L^-)&=&\frac{4 G_F}{\sqrt{2}}V^{\ast}_{ux}V_{cb}\Bigl[
C_1 \langle Q_1 \rangle + C_2 \langle Q_2 \rangle
\Bigl],
\end{eqnarray}
for $x=d,s$, $L^-=\pi^-\pi^0,~K^-\pi^0$ and $\displaystyle\langle Q_i \rangle \equiv \langle D^+ L^-| Q_i |\bar{B}^0 \rangle$. $Q_1$ and $Q_2$ are the current-current operators from the effective weak Hamiltonian and $C_{1,2}$ the corresponding Wilson coefficients. The factorization formula at leading power in $\Lambda_{\mathrm{QCD}}/m_b$ then reads
\begin{equation}
\mathcal{A}(\bar{B}\rightarrow D^+ L^-)=\frac{4 G_F}{\sqrt{2}} V^*_{ux}V_{cb} \, k^- F^{B\rightarrow D}_n \int^1_0 du [C_1 T_1(u) + C_2 T_2(u)] \phi_L (u,k),
\end{equation}
where $k^-=(m^2_B - m^2_D + k^2 + \sqrt{\lambda_{BD}(k^2)})/(2 m_B)$ and $\lambda_{ij}(k^2) = (m_i^2-m_j^2)^2 + k^4 - 2(m_i^2+m_j^2)k^2$ is the K\"all\'en function for the $BD$ pair. Moreover, we identify
\begin{eqnarray}
a_1(D^+L^-)&=&\int^1_0 du [C_1 T_1(u) + C_2 T_2(u)] \phi_L (u,k).
\label{eq:a1}
\end{eqnarray}
The functions $T_i$ in  Eq.~(\ref{eq:a1}) are  the corresponding hard scattering kernels. They are identical to the ones for two-body processes~\cite{Huber:2020pqb} and known to two loops in QCD~\cite{Huber:2016xod}. Moreover $\phi_L$ are the dimeson LCDAs~\cite{Polyakov:1998ze}. These latter objects are hardly constrained at all at the moment. They are non-local, non-perturbative objects with well-defined normalization and local limits in terms of time-like form factors~\cite{Huber:2020pqb,Polyakov:1998ze}.

The dimeson LCDA enjoys several expansions, the first one being in the eigenfunctions of their one-loop evolution kernel, $6u\bar u \, C^{3/2}_n(u-\bar u)$, where $C^{3/2}_n$ are the Gegenbauer polynomials,
\begin{eqnarray}
\phi_{L}(u,k)&=& 6 u \bar{u} \sum^{\infty}_{n=0} \alpha^{L}_n(k^2,\theta_\pi) C^{3/2}_n (u -\bar{u}) \, ,
\end{eqnarray}
followed by the expansion of the coefficients $\alpha^{L}_n(k^2,\theta_\pi)$ in partial waves, parameterized by the Legendre polynomials $P_\ell(\cos\theta_\pi)$,
\begin{eqnarray}
\alpha_n^{L}(k^2,\theta_\pi) = \sum_{\ell=0}^{n+1}
B^{L}_{n\ell}(k^2)\, P_\ell(\cos\theta_\pi) \, .
\label{eq:alphapipi}    
\end{eqnarray}
For $L=\pi\pi$, only even values for $n$ and odd values of $\ell$ contribute.

By convoluting the LCDAs $\phi_L$ with the hard scattering kernels $T_1$, $T_2$ as indicated in Eq.~(\ref{eq:a1}), ref.~\cite{Huber:2020pqb} determined the following expression for the square of the amplitude $a_1(D^+L^-)$ which shows explicitly the effect  of the NLO and NNLO QCD corrections

\vspace*{-5pt}

{\footnotesize{
\begin{eqnarray}
|a_1(D^+L^-)|^2 &=& |\alpha_0^L|^2 \,\Big\{
1.07_{\rm LO}\nonumber\\
&& + \big[0.053
- 0.026 \,\re\,\hat\alpha_1^L - 0.062\, \im\,\hat\alpha_1^L
+ 0.0047 \,\re\,\hat\alpha_2^L + 0.0034\, \im\,\hat\alpha_2^L
\big]_{\rm NLO}
\nonumber\\
&& + \big[0.029
- 0.091 \,\re\,\hat\alpha_1^L - 0.040\, \im\,\hat\alpha_1^L
+ 0.0036 \,\re\,\hat\alpha_2^L + 0.011\, \im\,\hat\alpha_2^L
\big]_{\rm NNLO}
\Big\}
\nonumber\\[2mm]
&=& 1.15 |\alpha_0^L|^2\,\Big\{
1
- 0.10\,\re\,\hat\alpha_1^L - 0.09\,\im\,\hat\alpha_1^L
+ 0.007\,\re\,\hat\alpha_2^L + 0.014\,\im\,\hat\alpha_2^L
\Big\} \, ,
\nonumber \\ 
&&
\end{eqnarray}
}}\normalsize
\noindent where $\hat\alpha_i^L\equiv \alpha_i^L/\alpha_0^L$. We can see that the corrections proportional to $\hat\alpha_1^L$ are of order $10\%$ with respect to those associated with $\hat\alpha_0^L$. Correspondingly, those related to $\hat\alpha_2^L$ lead to corrections of order $1\%$ in comparison. In addition, the NNLO effects are non-negligible, this effect is particularly noticeable for $\re \, \hat\alpha_1^L$ which is dominated by the NNLO corrections. This is most relevant for  $L=K\pi$ for which $\hat\alpha_1^{K\pi}\neq 0$.

\begin{figure}[htb]
\centering
\includegraphics[width=0.48\textwidth]{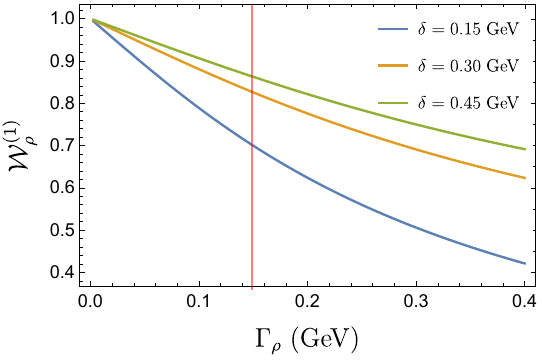}
\hspace{2mm}
\includegraphics[width=0.48\textwidth]{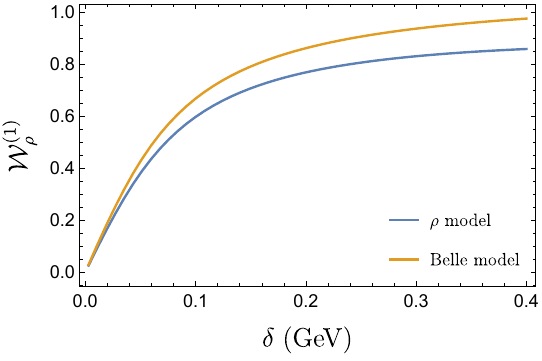}
	\caption{Finite-width and bin-size effects on $\mathcal{W}_\rho$. Left: Corrections to the narrow-width limit of the $\rho$ model used in \cite{Huber:2020pqb}. The vertical band displays the physical width $\Gamma_\rho = (149.1\pm0.8)$ MeV of the $\rho$. Right: $\rho$ model as outlined in the text and in~\cite{Huber:2020pqb}, and the Belle model \cite{Belle:2008xpe} as function of the bin size $\delta$.}
	\label{fig:finw}
\end{figure}

The coefficients $B^{L}_{nl}(k^2)$ can be extracted from data or can be modelled. For instance consider the following model that considers a sum over resonances \cite{Descotes-Genon:2019bud}
\begin{eqnarray}
B_{n0}^{M\pi}(s) &=& \sum_{R_0} \frac{m_{R_0}\, f_{R_0}\, g_{R_0M\pi}\, e^{i\varphi_{R_0}}}{\sqrt{2} [m_{R_0}^2 - s- i\,\sqrt{s}\, \Gamma_{R_0}(s)]}\, \alpha_n^{R_0}\ ,
\label{eq:Bn0}
\\
B_{n1}^{M\pi}(s) &=& \frac{\sqrt{\lambda_{M\pi}(s)}}{s} \sum_{R} \frac{m_{R}\, f_{R}\, g_{RM\pi}\, e^{i\varphi_{R}}}{\sqrt{2} [m_{R}^2 - s- i\,\sqrt{s}\,\Gamma_{R}(s)]}\, \alpha_n^{R}\ ,
\label{eq:Bn1}
\end{eqnarray}
which satisfies the narrow-width limit in the case of stable vector resonances such as the $\rho$ or the $K^{\ast}$ mesons. To investigate the corrections to the narrow width limit consider the decay rate integrated around a resonance 
\begin{eqnarray}
\Gamma_{[R]} \equiv \int_{(m_R-\delta)^2}^{(m_R+\delta)^2} ds\, \frac{d\Gamma(\bar B\to D^+M^-\pi^0)}{ds} = \sum_\ell \Gamma_{[R]}^{(\ell)}\ ,
\label{eq:GammaR}
\end{eqnarray}
where $\delta$ is the bin size and is taken large enough to capture the bulk of the contributions from the resonance $R$. In the narrow width limit we get $\Gamma^{(\ell)}_{[R],{\mathrm{NWL}}}=\Gamma(\bar{B}\rightarrow D^+ R^-)\mathcal{B}(R\rightarrow M \pi)$. To evaluate the effects of the finite width and bin sizes as well as corrections to the narrow width limit we define the function
\begin{eqnarray}
\mathcal{W}_R^{(\ell)} = \frac{\Gamma_{[R]}^{(\ell)}}{\Gamma_{{[R]}\,,{\rm NWL}}^{(\ell)}} \, ,    
\end{eqnarray}
whose behaviour for the $\rho$ model used in \cite{Huber:2020pqb}, which is largely based on Eq.~(\ref{eq:Bn1}), is shown in Fig.~\ref{fig:finw} and contrasted against the one mentioned in \cite{Belle:2008xpe}.


In order to assess the higher-order QCD effects as well as to have sensitivity to the coefficients $B^L_{n\ell}(k^2)$ we define the following ratio
\begin{equation}
\mathcal{R}_{M M'}[z_1,z_2;z_1',z_2'](k^2)\equiv\frac{\displaystyle\int_{z_1}^{z_2}\! dz \, \frac{d\Gamma(\bar B \to D^+ M^- \pi^0)}{dk^2 \, dz}}{\displaystyle\int_{z_1^\prime}^{z_2^\prime}\! dz \, \frac{d\Gamma(\bar B \to D^+ M^{\prime-} \pi^0)}{dk^2 \, dz}} = \frac{\displaystyle\int_{z_1}^{z_2}\! dz \, \left|a_1(D^+ M^-\pi^0)\right|^2}{\displaystyle\int_{z_1^\prime}^{z_2^\prime}\! dz \, \left|a_1(D^+ M^-\pi^0)\right|^2} \, , \label{eq:dalitzratios}   
\end{equation}
for $z=\cos\theta_{\pi}$. Using, Eq.~(\ref{eq:dalitzratios}) we find that the forward-backward asymmetry for the $D\pi\pi$ system vanishes,
\begin{eqnarray}
A^{\pi\pi}_{\rm FB}(k^2) = \mathcal{R}_{\pi\pi}[0,1;-1,1](k^2)-\mathcal{R}_{\pi\pi}[-1,0;-1,1](k^2) = 0 \, .   
\end{eqnarray}
Deviations from this result indicate corrections to the isospin limit. 

Assuming $P$-wave dominance and small $\alpha^{\pi\pi}_2$ it is possible to obtain the following result where the angular dependence factorizes from that on $k^2$, higher Gegenbauer moments and the NNLO QCD corrections (the latter two are encoded in the ${\cal G}_n$),
\begin{eqnarray}
&&\mathcal{R}_{\pi\pi}[z_1,z_2,z_1',z_2'](k^2) =\frac{\displaystyle I[z_1,z_2,P_1^2]}{I[z_1^\prime,z_2^\prime,P_1^2]} \nonumber \\[0.3em]
&&+ \frac{\displaystyle I[z_1,z_2, P_1 \, P_3] \, I[z_1^\prime,z_2^\prime,P_1^2] - I[z_1^\prime,z_2^\prime, P_1 \, P_3] \, I[z_1,z_2,P_1^2]}{\left(I[z_1^\prime,z_2^\prime,P_1^2]\right)^2} \nonumber \\[0.3em]
&&\; \times \frac{\displaystyle 2 \, {\rm{Re}}\left(B^{\pi\pi}_{01}(k^2) \, B^{\pi\pi \, \ast}_{23}(k^2) \, {\cal G}_0(\mu_b) \, {\cal G}_2^\ast(\mu_b) \right) + 2 \, {\rm{Re}}\left(B^{\pi\pi}_{21}(k^2) \, B^{\pi\pi \, \ast}_{23}(k^2) \right) |{\cal G}_2(\mu_b)|^2}{\displaystyle \left|B^{\pi\pi}_{01}(k^2) \, {\cal G}_0(\mu_b) + B^{\pi\pi}_{21}(k^2) \, {\cal G}_2(\mu_b) \right|^2} \, . \nonumber \\ && \label{eq:dipionPwavedom}    
\end{eqnarray}

\begin{figure}[htb]
\centering
\includegraphics[width=0.75\textwidth]{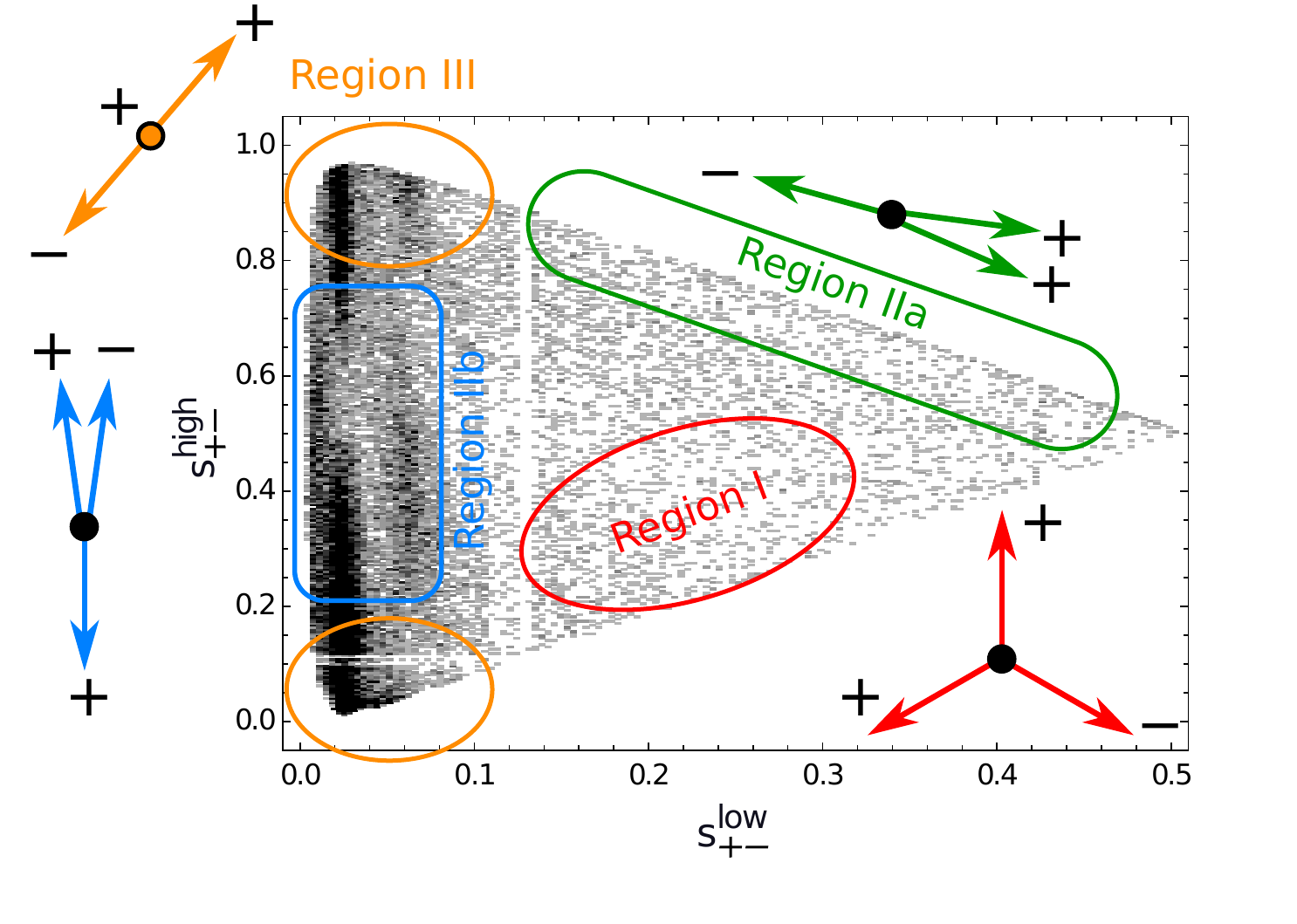}

\vspace*{-20pt}

	\caption{Kinematical regions in the Dalitz plot for the decay $B^+\rightarrow \pi^+\pi^-\pi^+$. Graphic courtesy of Javier Virto.\label{fig:regions}}	
\end{figure}

\subsubsection{Three light final states}

\begin{figure}[htb]
\begin{center}
\includegraphics[width=5cm]{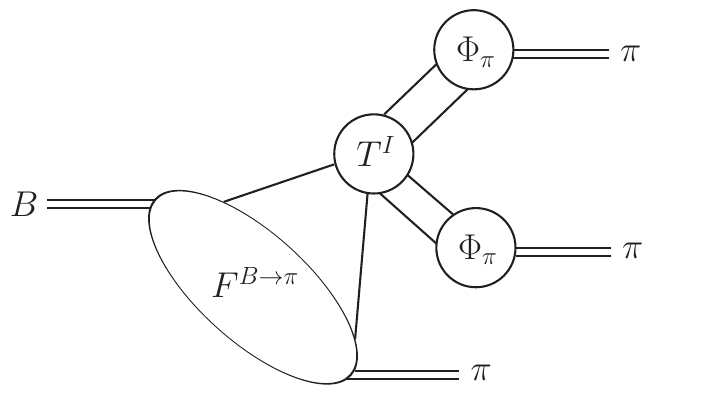} 
\hspace{1cm}
\raisebox{15mm}{+}
\hspace{1cm}
\includegraphics[width=5cm]{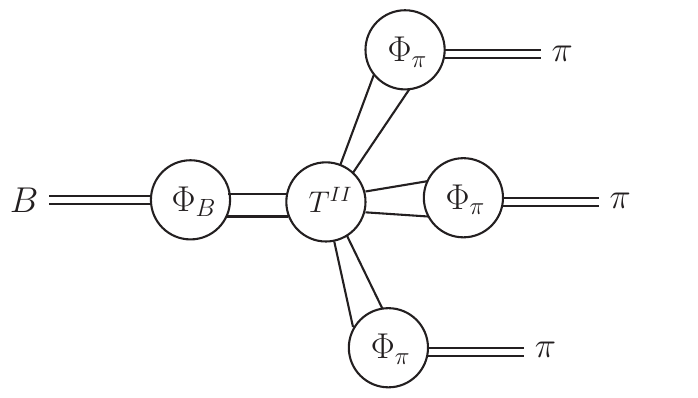}
\end{center}
\caption{Factorization formula in the center (region I). Figure from~\cite{Krankl:2015fha}.\label{fig:FactCenter}}
\end{figure}

If the three final-state mesons are light then it is convenient to divide the phase space in different regions depending on the invariant masses of the particle pairs in the final state \cite{Krankl:2015fha}. Three regions can be identified and are shown in Fig. \ref{fig:regions}. The region I corresponds to the central area of the Dalitz plot where all the invariant masses are approximately the same and equal to $\approx 1/\sqrt{3}m_B$. The region II is characterized by one small invariant mass and two large ones, in this situation two pions are collinear and recoil against the third one. Finally, in the region III, two invariant masses are small and corresponds to the situation where two pions are fast and back to back whereas the third one is soft.  The factorization rules for each region are different. For instance, in the region I we have 
\begin{equation} 
\langle \pi^+\pi^-\pi^+ | {\cal O}_i |B^+ \rangle_{s_{ij} \sim 1/3} = T_i^I \otimes F^{B\to \pi} \otimes \Phi_{\pi} \otimes \Phi_{\pi}
+ T_i^{II} \otimes \Phi_B \otimes \Phi_{\pi} \otimes \Phi_{\pi} \otimes \Phi_{\pi} \ .
\label{eq:FactCenter}
\end{equation} 
This formula is pictorially shown in Fig.~\ref{fig:FactCenter}. It is found that at tree level (${\mathcal{O}}(\alpha_s)$) all the convolutions are finite. Although higher orders in QCD have not been calculated yet, the ${\mathcal{O}}(\alpha^2_s)$ corrections are expected to be of order $10\%$ with respect to the leading color-allowed amplitude~\cite{Krankl:2015fha}, similar to $B \to \pi\pi$~\cite{Beneke:2015wfa}.

\begin{figure}[htb]
\begin{center}
\includegraphics[width=5cm,height=3.3cm]{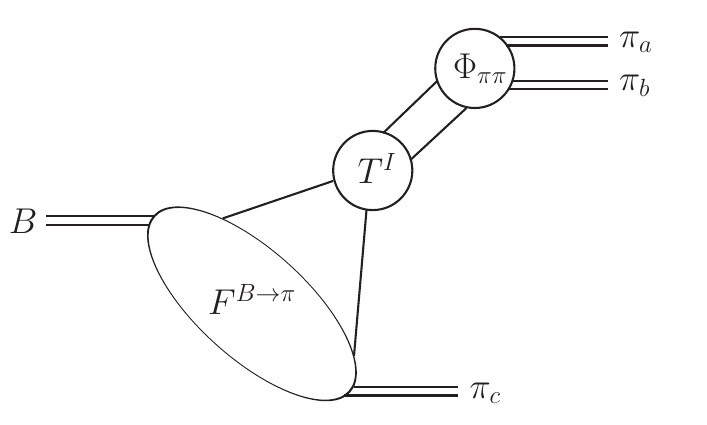} 
\hspace{-5mm}
\raisebox{12mm}{+}
\hspace{5mm}
\includegraphics[width=5cm,height=3.3cm]{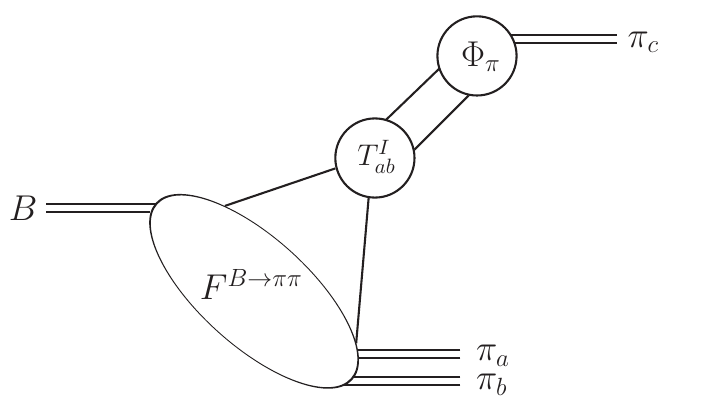}
\hspace{-5mm}
\raisebox{12mm}{+}
\hspace{5mm}
\includegraphics[width=5cm,height=3.0cm]{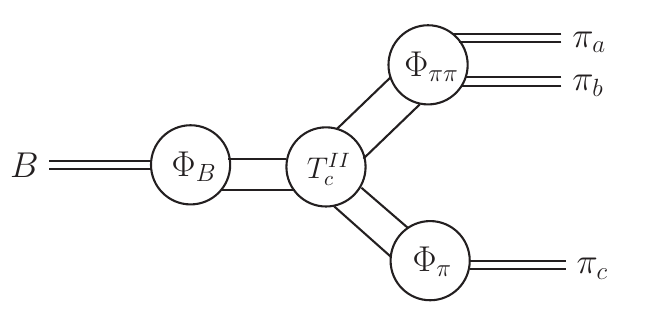}
\end{center}
\caption{Factorization formula for the edges of the Dalitz plot (region II).  Figure from~\cite{Krankl:2015fha}.\label{fig:FactEdge}}
\end{figure}

In the region II, which lies in the edges of the Dalitz plot, the relevant factorization formula is (see Fig.~\ref{fig:FactEdge})
\begin{eqnarray} 
\langle\pi^a\pi^b\pi^c|{\cal O}_i|B\rangle_{s_{ab} \ll 1} &=& T_c^I \otimes F^{B\to \pi^c} \otimes \Phi_{\pi^a\pi^b}
+ T_{ab}^I \otimes F^{B\to \pi^a\pi^b} \otimes \Phi_{\pi^c}\nonumber \\[2mm]
&+& T^{II} \otimes \Phi_B \otimes \Phi_{\pi^c} \otimes \Phi_{\pi^a\pi^b}\, .
\label{s23small}
\end{eqnarray}
Here the three-body decays resemble the two-body processes and resonant channels can be produced. In order to describe this region it is necessary to introduce $2\pi$ LCDAs~\cite{Polyakov:1998ze} as well as $B\rightarrow \pi\pi$ form factors which can be extracted from decays such as $B\rightarrow \pi \pi \ell\nu$ \cite{Faller:2013dwa}.

\subsection{Charmless $B$ decays at Belle II}

Among the six observables of the UT, the angle $\phi_2 ({\rm or} ~\alpha)\equiv\arg(-V_{td}V_{tb}^*/V_{ud}V_{ub}^*)$, is the least precisely measured, $\phi_2 = (85.2^{+4.8}_{-4.3})^o$~\cite{HFLAV:2022esi}.  One approach to measure $\phi_2$ is to measure the time-dependent decay-rate
asymmetry between $\bar{B}^0$ and $B^0$ mesons that decay to $\pi^+\pi^-$ final states. The above asymmetry gets contribution from tree-level $b\to u$ processes but is affected due to the presence of $b \to d$ penguin contributions. The penguin contributions can be disentangled using the $B \to \pi\pi$ isospin relations ~\cite{Gronau:1990ka}
\begin{eqnarray}
A^{+0}= \frac{1}{\sqrt(2)}A^{+-} + A^{00}, && {\bar A}^{-0}= \frac{1}{\sqrt(2)}{\bar A}^{+-} + \bar{A}^{00}. \end{eqnarray}
Here, $A^{ij}$ and $\bar A^{ij}$ are amplitudes for the decays $B \to \pi^i\pi^j$
and $\bar B \to \pi^i\pi^j$, respectively. Here, $B$ and $\pi$ indicate charged or neutral bottom-mesons and pions, respectively, while $i$
and $j$ refer to electric charge. 
The above isospin relations require precise measurements of branching fraction $(\cal B)$ and time dependent $CP$ asymmetry for the $B^0 \to \pi^+\pi^-$ decay, together with measurements of $\cal B$ and the direct $CP$ asymmetry $({\cal A}_{CP})$ for $B^+ \to \pi^+\pi^0$ and $B^0 \to \pi^0\pi^0$ decays.  Among the three $B \to \pi\pi$ decays, the observables for $B^0 \to \pi^0\pi^0$ are the least well determined.  Belle II is the only experiment to competitively study this channel. \\
The other approach which is the most promising way to determine $\phi_2$ is based on the combined analysis of the decays $B^+ \to \rho^+\rho^0$ , $B^0 \to \rho^+\rho^-$, $B^0 \to \rho^0\rho^0$ because $\rho\rho$ channel has smaller penguin pollution compared to $B \to \pi\pi$~\cite{Charles:2017evz}. Using efficient reconstruction of low-energy $\pi^0$, improved measurements in $B^+ \to \rho^+\rho^0$ and $B^0 \to \rho^+\rho^-$ decays are unique to Belle II.

The challenge in analysing these fully hadronic final states lies in the large amount of $e^+ e^- \to  q\bar q$ continuum background which is $O(10^6)$ higher than signal. Binary-decision-tree classifier output $C$ are used to discriminate between signal and continuum events using event topology, kinematic, and decay-length information. Signal efficiency is maximized with loose cuts and include the output $C$ in the fit to gain signal-to-background discrimination. The determination of the signal yields is mainly based on the energy difference between the $B$ candidate and the beam energy,
$\Delta E = E^{*}_B - E^{*}_{\rm beam}$ , and the beam-constrained mass $M_{bc} = \sqrt{E^{*}_{\rm beam}/c^2 - (p^{*}_{B}/c)^2}$, where $E^{*}_B$ and $p^{*}_B$ are the energy and momentum of the $B$ candidate, respectively, and $E^{*}_{\rm beam}$ is the beam energy. Here $*$ indicates that the quantity is evaluated in the center-of-mass frame.

The Belle II collaboration reports the measurements of  $B^0 \to \pi^+\pi^-$ and $B^+ \to \pi^+\pi^0$ decays using full LS1 data set corresponding to $362~\rm fb^{-1}$ of integrated luminosity recorded at the $\Upsilon(4S)$ resonance. Signal yields are determined with an extended two-dimensional maximum likelihood fit of the unbinned energy-difference $\Delta E$ and transformed continuum suppression classifier ($C^{'}$) distributions. The results are competitive with world's best results. The dominant uncertainty for ${\cal B}(B\to \pi^+\pi^0)$ comes from the $\pi^0$ efficiency. The first measurement of $B^0 \to \pi^0\pi^0$ at Belle II using $189~\rm fb^{-1}$ of data is also reported~\cite{Belle-II:2023cbc}.  This decay is both CKM- and colour-suppressed, and has only photons in the final state, making it experimentally challenging to measure. The result obtained from a fit to $M_{bc}$, $\Delta E$, and $C^{'}$, achieves Belle’s precision despite using a dataset that is only one third of Belle sample size. This is due to the dedicated photon selection and continuum suppression studies that yield a much higher $\pi^0$ efficiency. The measured values of ${\cal B}$ and ${\cal A}_{CP}$ are reported to be 
\begin{eqnarray}
{\cal B}(B^0 \to \pi^+\pi^-)  &=&  (5.83 \pm 0.22 \pm 0.17) \times 10^{-6},\\
{\cal B}(B^+ \to \pi^+\pi^0)  &=&  (5.10 \pm 0.29 \pm 0.32) \times 10^{-6},\\
{\cal A}_{CP}(B^+ \to \pi^+\pi^0)  &=&  -0.081 \pm 0.54 \pm 0.008,\\
{\cal B}(B^0 \to \pi^0\pi^0)  &=&  (1.38 \pm 0.27 \pm 0.22) \times 10^{-6},\\
{\cal A}_{CP}(B^0 \to \pi^0\pi^0)  &=&  0.14 \pm 0.46 \pm 0.07.
\label{eq:belleii_btocharmless1}
\end{eqnarray}

 Belle II reported measurement of $\cal B$, ${\cal A}_{CP}$, and the fraction of longitudinal polarized decays $f_L$ of $B \to \rho \rho$ decays reconstructed in $189~\rm fb^{-1}$ of Belle II data. The measurements require a complex angular analysis. Signal yields are determined with likelihood fits of the unbinned distributions of $\Delta E$, $C^{'}$, the dipion masses, and the helicity angle of the $\rho$ candidates. The preliminary Belle II results of $B^0 \to \rho^+ \rho^-$ and $B^0 \to \rho^+ \rho^0$ decays using $189~\rm fb^{-1}$ of data~\cite{Belle-II:2022ihd, Belle-II:2022ksf} are on par with the best performances from Belle~\cite{Belle:2015xfb, Belle:2003lsm} and BaBar~\cite{BaBar:2007cku, BaBar:2009rmk}. Major systematic uncertainty
from data-MC mismodelling needs improvement. Results are reported to be

\begin{eqnarray}
{\cal B}(B^0 \to \rho^+\rho^-)  &=&  (26.7 \pm 2.8 \pm 2.8) \times 10^{-6},\\
{f_L}(B^0 \to \rho^+\rho^-)  &=&  0.956 \pm 0.035 \pm 0.033,\\
{\cal B}(B^+ \to \rho^+\rho^0)  &=&  (23.2^{+2.2}_{-2.1} \pm 2.7) \times 10^{-6},\\
{f_L}(B^+ \to \rho^+\rho^0)  &=&  0.943^{+0.035}_{-0.033} \pm 0.027,\\
{\cal A}_{CP}(B^+ \to \rho^+\rho^0) &=&  - 0.069 \pm 0.068 \pm 0.060 \, .
\label{eq:belleii_btocharmless2}
\end{eqnarray}

The so-called $K\pi$ puzzle is a long-standing anomaly associated with the significant difference between direct $CP$-violating asymmetries observed in $B^0 \to K^+\pi^-$ and $B^+ \to K^+ \pi^0$ decays~\cite{HFLAV:2022esi}. The asymmetries are expected to be equal at the leading order, as the two decays differ only in the spectator quark. To check whether difference between direct $CP$-violating asymmetries is due to strong dynamic effects or due to new physics, a more precise test of the isospin-sum rule~\cite{Gronau:2005kz} is suggested where 
\begin{equation}
I_{K\pi}= {\cal A}_{K^+\pi^-} + {\cal A}_{K^0\pi^+}.\frac{{\cal B}(K^0\pi^+)}{{\cal B}(K^+\pi^-)}\frac{\tau_{B^0}}{\tau_{B^+}} - 2{\cal A}_{K^+\pi^0}.\frac{{\cal B}(K^+\pi^0)}{{\cal B}(K^+\pi^-)}\frac{\tau_{B^0}}{\tau_{B^+}} - 2{\cal A}_{K^0\pi^0}.\frac{{\cal B}(K^0\pi^0)}{{\cal B}(K^+\pi^-)}.\end{equation}
SM predicts $I_{K\pi} \simeq 0$ under isospin symmetry and assuming no electroweak penguin (EWP) contributions, with an uncertainty of $O(1\%)$.
To precisely test the above sum rule, it is necessary to precisely measure $\cal B$ and ${\cal A}_{CP}$ of $B$ decays to all four final states $K^+\pi^-$, $K^0 \pi^+$, $K^+ \pi^0$ and $K^0 \pi^0$. Belle II has the unique capability of studying jointly and within a consistent experimental environment, all the above final states. Specifically the experiment will be unique in measuring ${\cal A}_{CP}$ in $B^0 \to  K^0 \pi^0$ decays, the input that limits the precision of the isospin sum rule~\cite{pdg}.

\begin{figure}[htb]
\includegraphics[width=1
\linewidth]{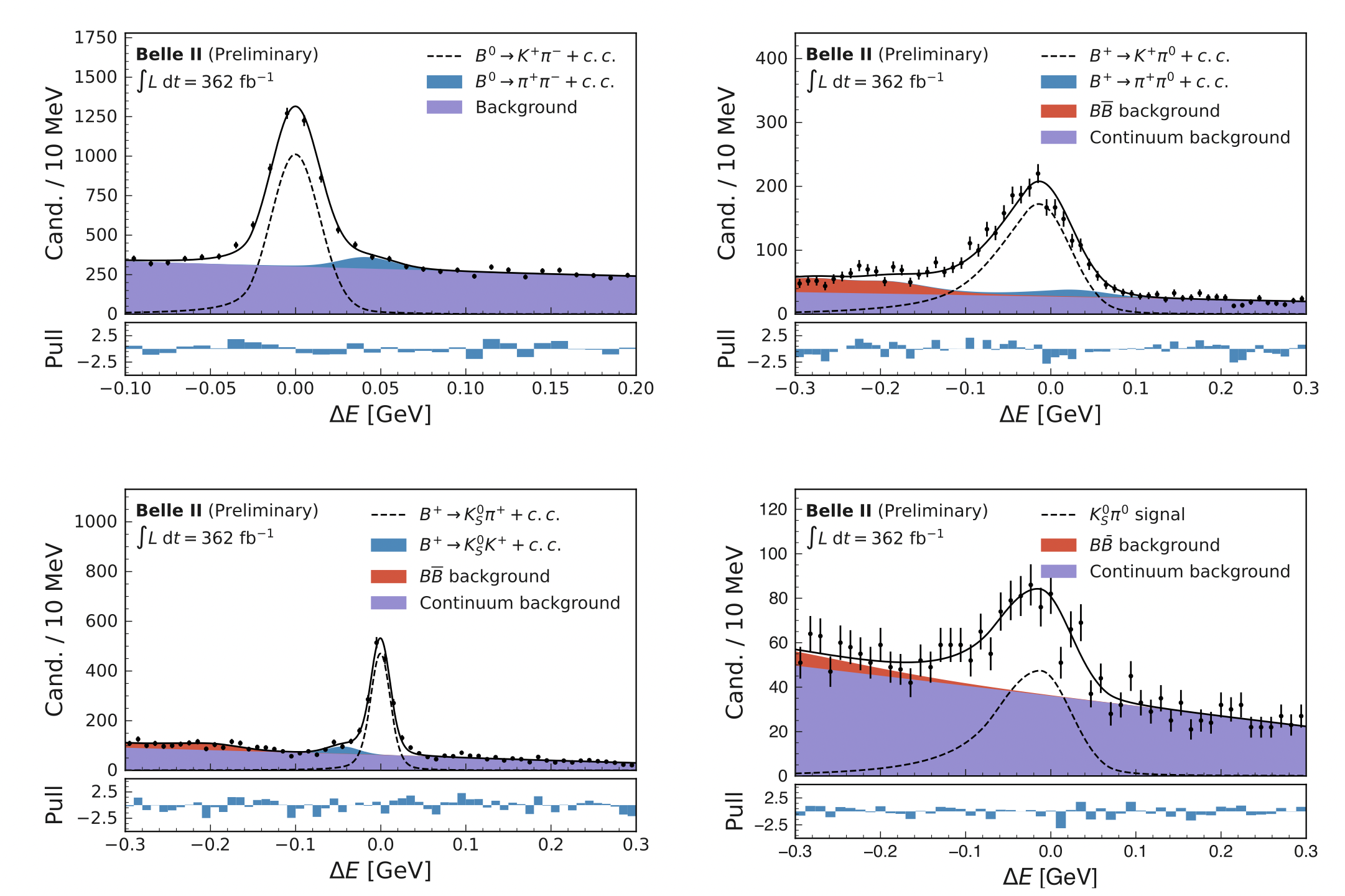}
\caption{$\Delta E$ distributions of $B^0 \to K^+\pi^-$ (upper left), $B^+ \to K^+\pi^0$ (upper right), $B^0 \to K_S^0\pi^+$ (lower left) and $B^0 \to K_S^0\pi^0$ (lower right) decays.}
\label{fig:belleii_sumrule} 
\end{figure}

Belle II reported measurement of $\cal B$, ${\cal A}_{CP}$ using $362~\rm fb^{-1}$ of data. The analyses of the various decays follow a similar strategy, with common selections applied to the final states particles. $B$ candidates are required to satisfy $5.272 < M_{bc} < 5.288 ~{\rm GeV}/c^2$, $|\Delta E| < 0.3~\rm GeV$, and a loose requirement of $C$ that suppresses $90-99\%$ of continuum background. A fit is performed on the $\Delta E - C'$ distribution, where the flavour tagging algorithm~\cite{Belle-II:2021zvj} is employed to determine the flavour of the $B$ candidate in $B^0 \to K^0_{S} \pi^0$ decay due to absence of primary charged particles. The $\Delta E$ distributions are shown in Fig.~\ref{fig:belleii_sumrule}. 

The measured branching fractions and $CP$ asymmetries, as  well as the Belle II isospin sum-rule calculated using these measurements, are summarised below.
\begin{eqnarray}
{\cal B}(B^0 \to K^+\pi^-)  &=&  (20.67 \pm 0.37 \pm 0.62) \times 10^{-6},\\
{\cal A}_{CP}(B^0 \to K^+\pi^-)  &=&  -0.072 \pm 0.019 \pm 0.007,\\
{\cal B}(B^+ \to K^+\pi^0)  &=&  (14.21 \pm 0.38 \pm 0.84) \times 10^{-6},\\
{\cal A}_{CP}(B^+ \to K^+\pi^0)  &=&  0.013 \pm 0.027 \pm 0.005\\
{\cal B}(B^+ \to K^0\pi^+)  &=&  (24.39 \pm 0.71 \pm 0.86) \times 10^{-6},\\
{\cal A}_{CP}(B^+ \to K^0\pi^+)  &=&  0.046 \pm 0.029 \pm 0.007\\
{\cal B}(B^0 \to K^0\pi^0)  &=&  (10.50 \pm 0.62 \pm 0.65) \times 10^{-6},\\
{\cal A}_{CP}(B^0 \to K^0\pi^0)  &=&  -0.01 \pm 0.12 \pm 0.05,\\
{I}_{K\pi}  &=&  -0.03 \pm 0.13 \pm 0.05.
\label{eq:belleii_btocharmless3}
\end{eqnarray}

They agree with the world averages and have competitive precisions. In particular, the time-integrated and time-dependent results of $B^0 \to K^0_{S} \pi^0$ are combined to achieve the world’s best result for ${\cal A}_{CP}$, and consequentially for $I_{K\pi}$ a competitive precision that is limited by the statistical uncertainty.

\subsection{Charmless $B$ decays at LHCb}

$B$-hadron decays without charm or charmonium contributions in their final states, are characterized by transitions of the type $b \to u (s)$ or $b \to d$. These, denominated as charmless $B$ decays, provide a rich environment for $CP$ violation studies as dominant tree-level and Penguin diagrams contribute in the same order of magnitude. Multi-body decays are dominated  by rich resonant structures that can give rise to $CP$ violation signatures localized in regions of the phase space. Their study is  particularly interesting as can shed light in the understanding of the $B$-hadron dynamics and offers the possibility to search for new sources of $CP$-violation~\cite{BediagaGobel}. The role of short- and long-distance contribution to the generation of strong-phase differences needed for direct $CP$ violation to occur can be investigated in three-body $B$ decays.

\begin{figure}[htb]
\includegraphics[width=0.5
\linewidth]{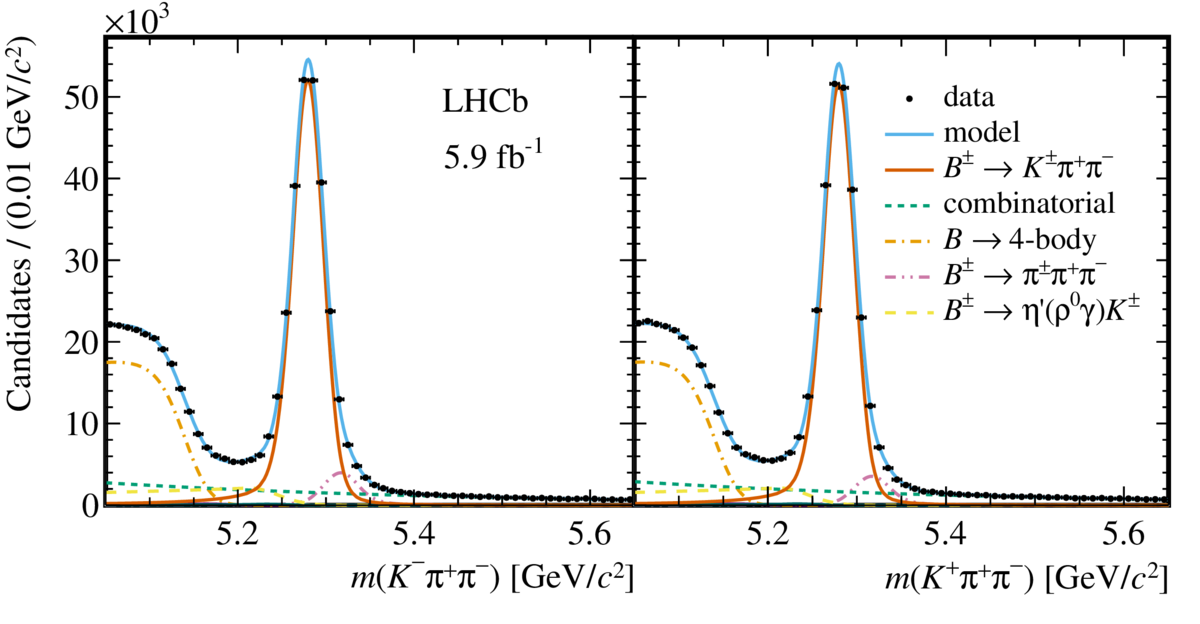}
\includegraphics[width=0.48
\linewidth]{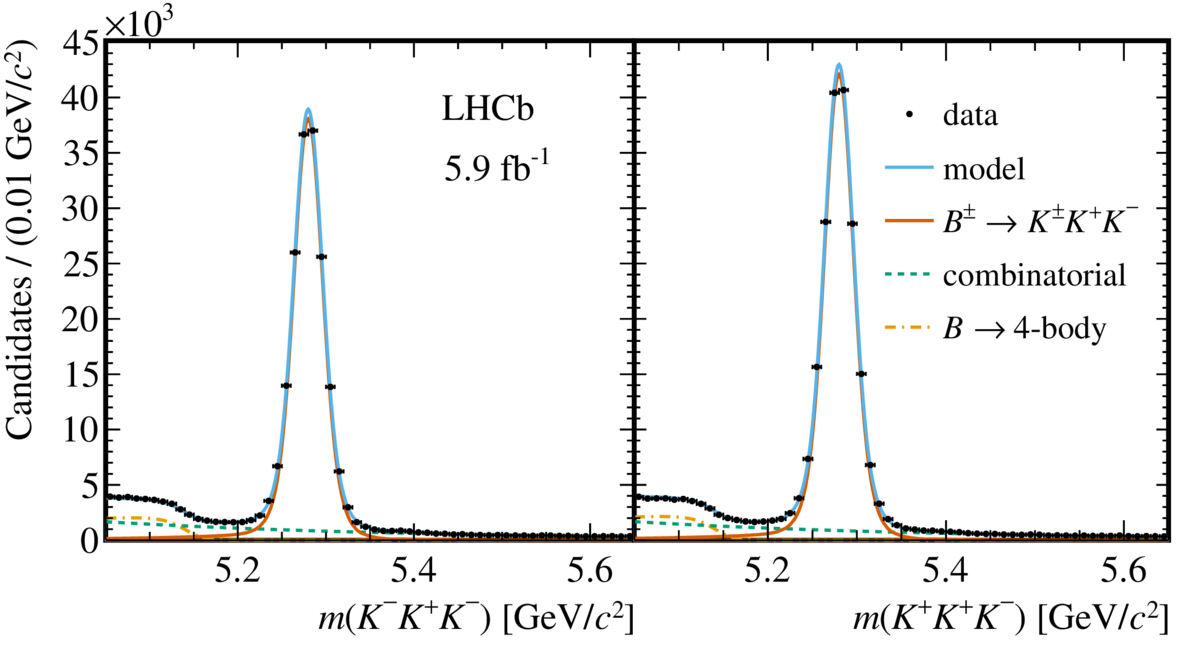}
\includegraphics[width=0.5
\linewidth]{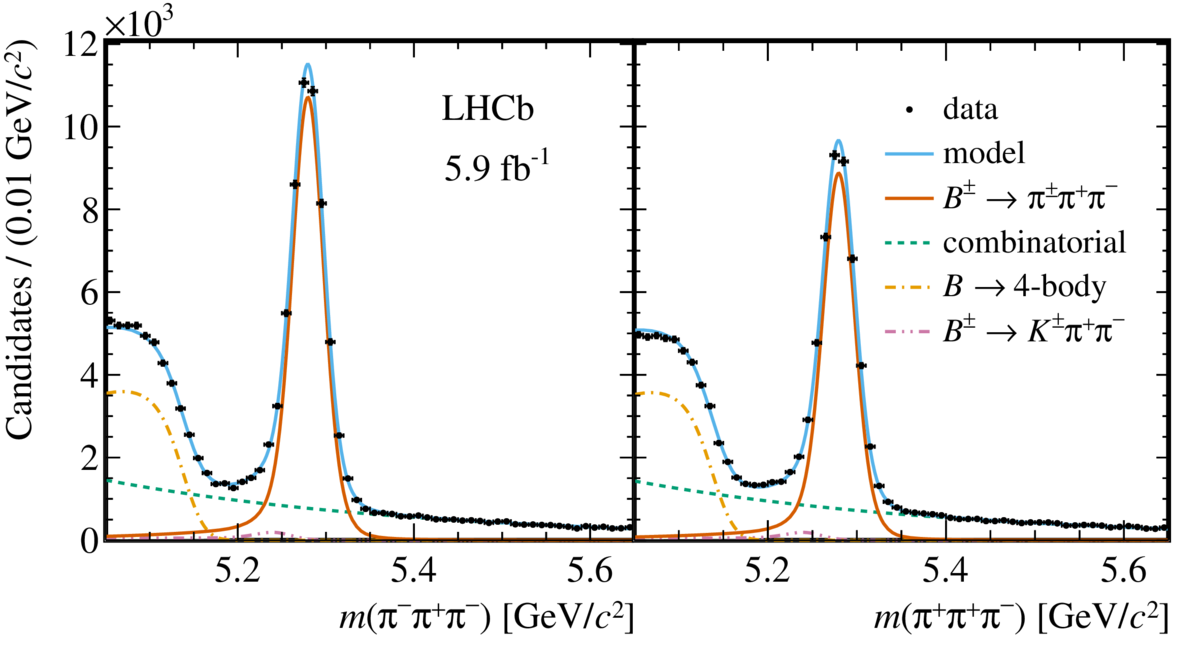}
\includegraphics[width=0.5
\linewidth]{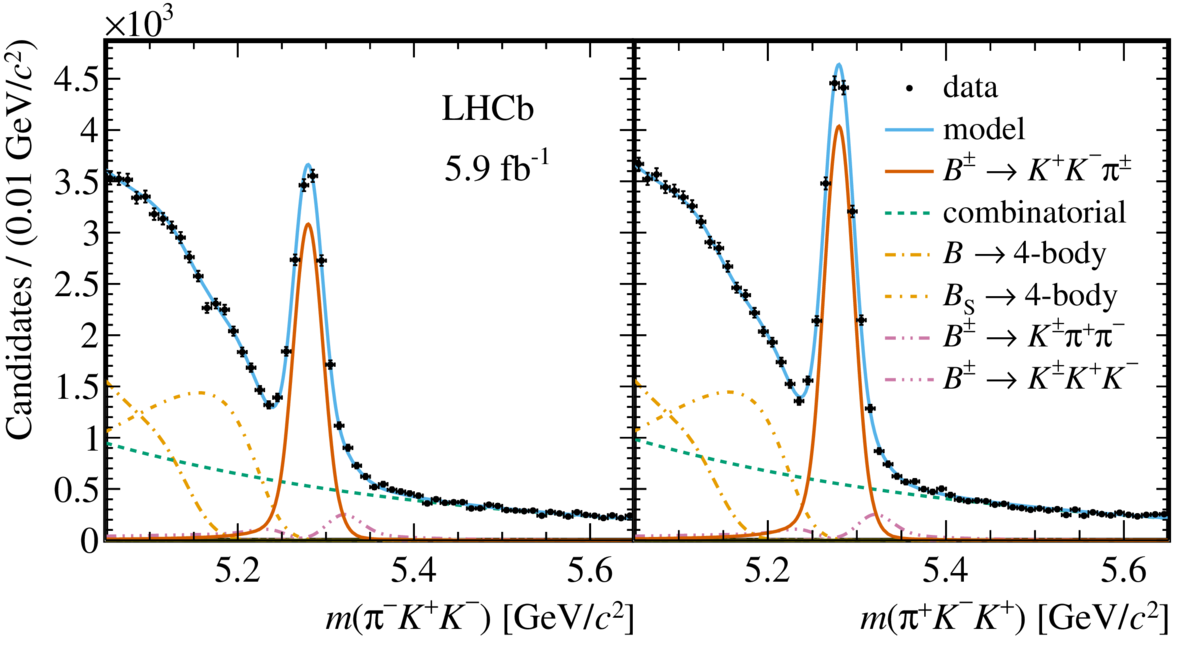}

\caption{Unbinned maximum likelihood fit to the invariant mass distribution of (a) $B^\pm \to K^\pm \pi^+\pi^-$, (b) $B^\pm \to K^\pm K^+K^-$, (c) $B^\pm \to \pi^\pm \pi^+\pi^-$ and (d) $B^\pm \to K^\pm K^+K^-$. In each panel (left )$B^-$ and (right) $B^+$ candidates.}
\label{fig:fig3.1} 
\end{figure}

The LHCb collaboration reports the updated measurements of $CP$ asymmetries in charmless three-body decays of $B^\pm$ mesons using the full Run 2 dataset corresponding to an integrated luminosity $5.9$ fb$^{-1}$ collected at the center-of-mass energy of 13 \rm{TeV} in 2015 to 2018~\cite{LHCb:3h-1}~\cite{LHCb:3h-2}. Four charmless $B$ decays to three charged pseudoscalar particles are analysed: $B^\pm \to K^\pm \pi^+ \pi^-$, $B^\pm \to K^\pm K^+ K^-$, $B^\pm \to \pi^\pm K^+ K^-$ and $B^\pm \to \pi^\pm \pi^+ \pi^-$. Given the similar topology shared among the decays modes, common selection strategies are applied based on simulation studies and data-driven techniques. A simultaneous extended maximum-likelihood fit is applied to $B^+$ and $B^-$ invariant mass distribution in order to extract the signal yields and thus, to calculate raw asymmetries. The latter defined as 

\begin{equation}
    A_{raw} = \frac{N^- - N^+}{N^- + N^+},
\end{equation}
 where $N$ are the observed signal yields. The probability density function (PDF) used to perform the one-dimensional mass fit is composed by the sum of functions that parameterize signal and backgrounds events. The fit to the invariant mass distribution for the four channels can be seen in Fig.~\ref{fig:fig3.1}.

 To extract the physical $CP$ asymmetry, corrections due to experimental effects need to be applied to the $A_{raw}$. The selection efficiencies and raw asymmetries do not distribute uniformly across the phase space (or Dalitz plot)~\cite{}. Therefore a two-dimensional efficiency models in the square Dalitz plot variables are constructed separately for $B^+$ and $B^-$. The phase-space integrated $CP$ asymmetry, denominated as $A_{CP}$, is obtained by taking into account the corrections for selection efficiency effects and the production asymmetry $A_P$. The efficiency-corrected raw asymmetry is denoted as $A_{raw}^{corr}$, and the production asymmetry for $B^\pm$ is obtained using the control channel $B^\pm \to J/\psi(\to \mu^+\mu^-)K^\pm$. Its reported value is $A_p = -0.0070 \pm 0.0008^{+0.0007}_{-0.0008} \pm 0.0030$, where the first uncertainty is statistical, the second systematic and the last one due to the $CP$ asymmetry of $B^\pm \to J/\psi K^\pm$ decays~\cite{LHCb:3h-1}.
 
 The physical asymmetry due to $CP$-violation is then expressed as:

 \begin{equation}
     A_{CP} = \frac{A_{raw}^{corr} - A_P}{1-A_{raw}^{corr}A_P}
 \end{equation}

The measured values of the $CP$ asymmetries for the four channels $B^\pm \to h^\pm h^{'+}h^{'-}$ are reported to be~\cite{LHCb:3h-1} 

\begin{center}
$A_{CP}(B^\pm \to K^\pm \pi^+\pi^-) = +0.011 \pm 0.002 \pm 0.003 \pm 0.003$, \\
$A_{CP}(B^\pm \to K^\pm K^+K^-) = -0.037 \pm 0.002 \pm 0.002 \pm 0.003$, \\
$A_{CP}(B^\pm \to \pi^\pm \pi^+\pi^-) = +0.080 \pm 0.004 \pm 0.003 \pm 0.003$, \\
$A_{CP}(B^\pm \to \pi^\pm K^+K^-) = -0.114 \pm 0.007 \pm 0.003 \pm 0.003$,
\end{center}
where the first uncertainty is statistical, the second systematic and the third is due to the limited knowledge of the $CP$ asymmetry of the $B^\pm \to J/\psi K^\pm$ control mode~\cite{pdg}. These results show a significant inclusive $CP$ asymmetry for the latter three channels, being the first observations for the last two. For $B^\pm \to K^\pm \pi^+\pi^-$ no $CP$ violation is confirmed.

\begin{figure}[htb]
\includegraphics[width=0.5
\linewidth]{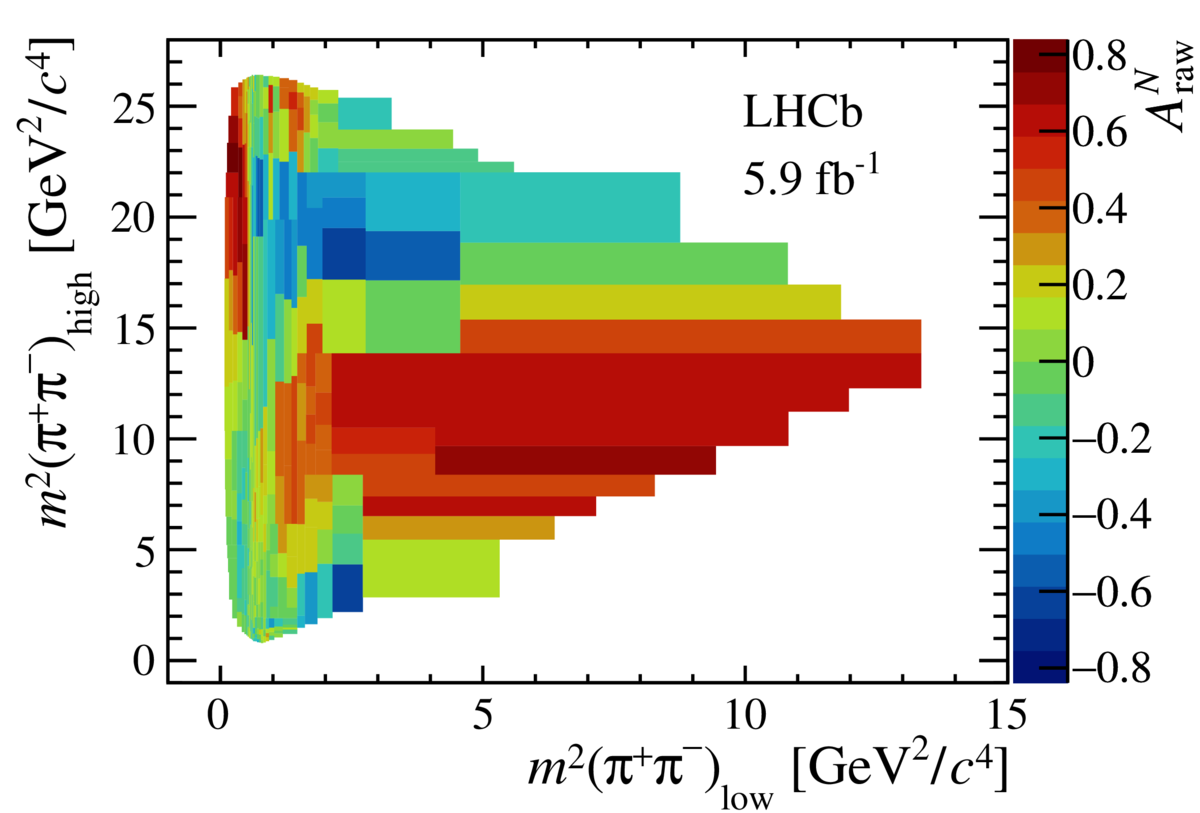}
\includegraphics[width=0.5
\linewidth]{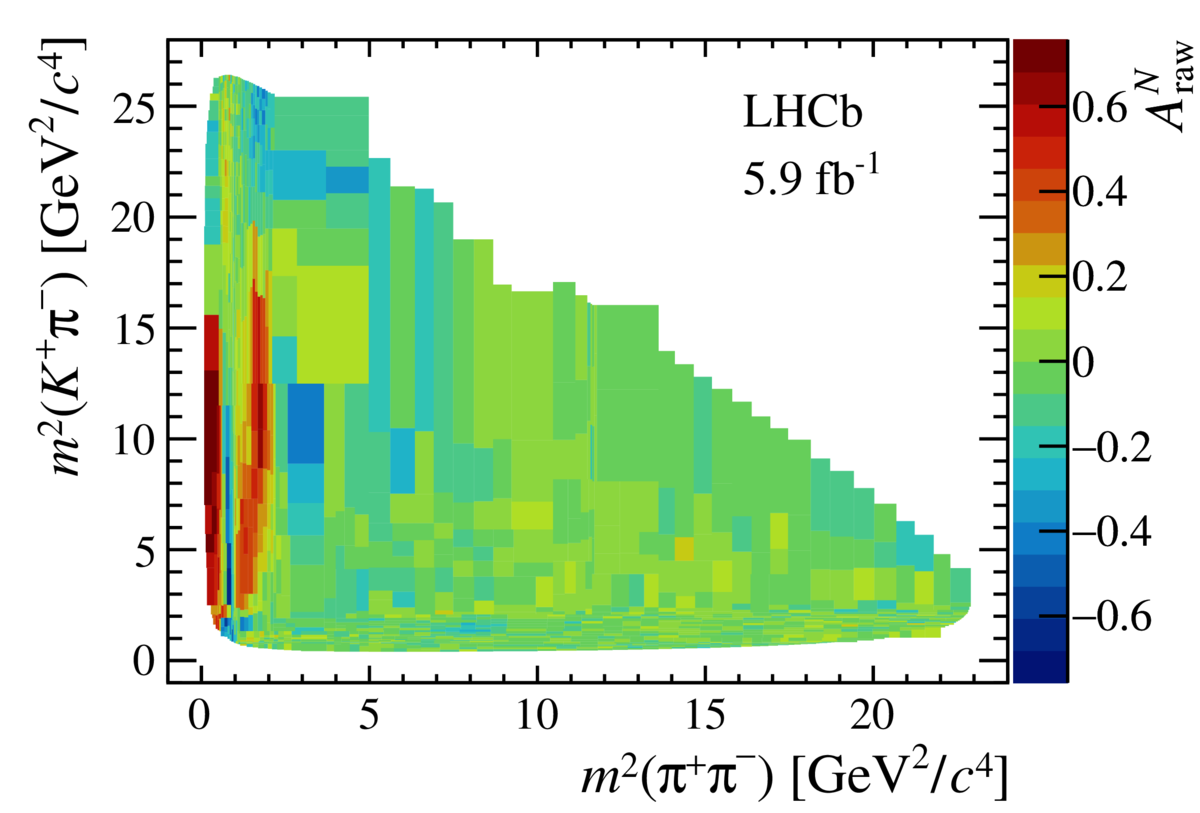}
\includegraphics[width=0.5
\linewidth]{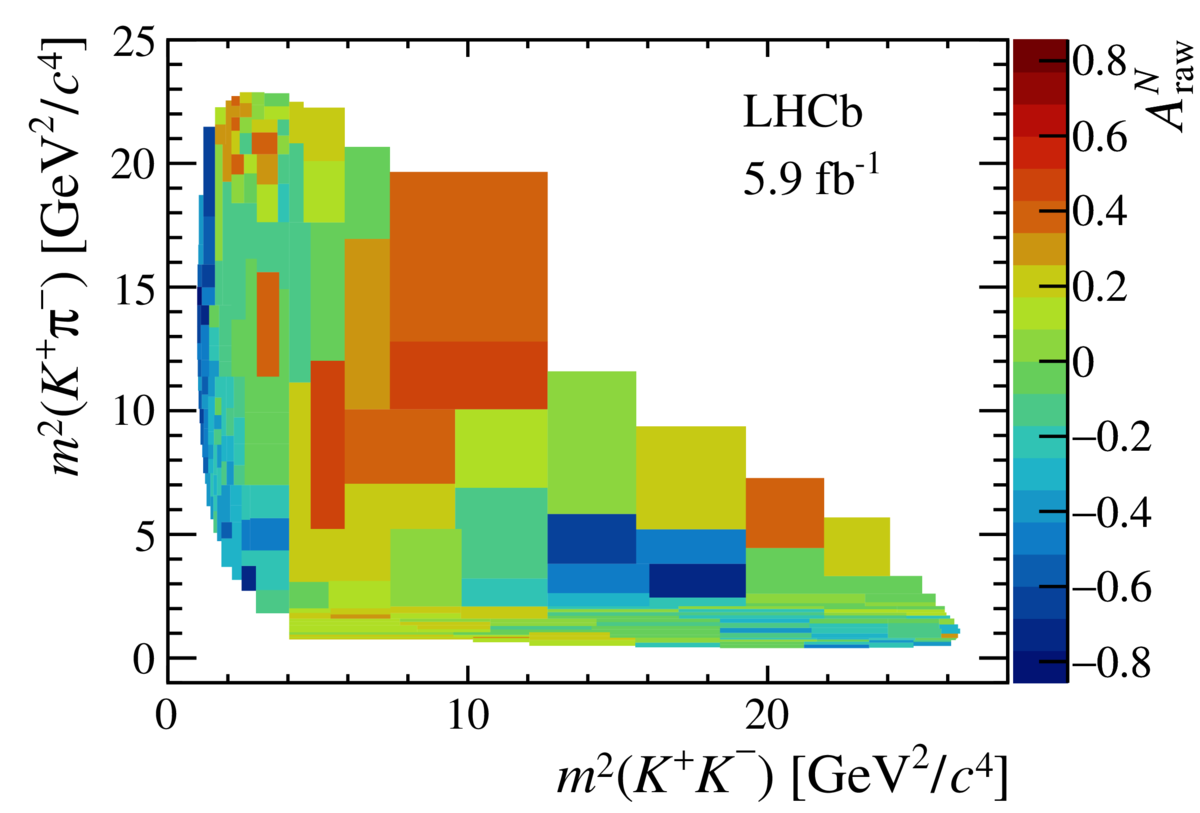}
\includegraphics[width=0.5
\linewidth]{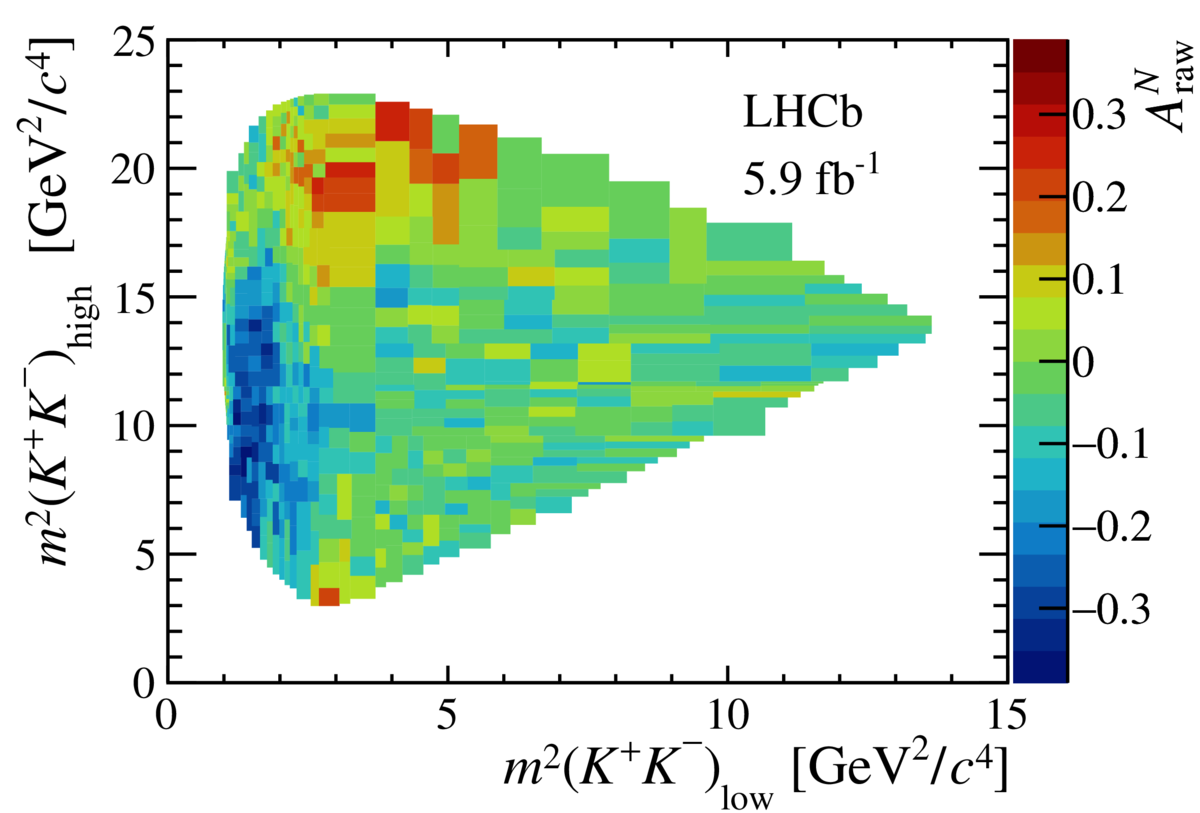}

\caption{Asymmetries distribution in bins of Dalitz plot. (a) $B^\pm \to \pi^\pm \pi^+\pi^-$, (b) $B^\pm \to K^\pm \pi^+\pi^-$, (c)  $B^\pm \to \pi^\pm K^+K^-$ and (d) $B^\pm \to K^\pm K^+K^-$.
\label{fig:fig3.2}}
\end{figure}

The two-dimensional phase space, the so called Dalitz plot, is expressed as function of the squared invariant masses of two of the three possible particle pairs.  Any non-uniformity of event in the phase space reflects directly the dynamics underlying the decay process, moreover the study of the resonant and nonresonant components allows to inspect for $CP$ violation effects. In order to study localized asymmetries, the $A_{CP}$ is constructed in bins of Dalitz plot. A rich pattern was obtained evidencing large localized asymmetries as result of the interference of the resonant intermediate states. Large asymmetries are  observed in the $\pi\pi \to KK$ rescattering region, as observed in the amplitude analyses studies~\cite{LHCb:AmAn3p-1, LHCb:AmAn3p-2} \cite{LHCb:AmAnkkpi}. The asymmetries distribution for the four decays modes are shown in Fig.~\ref{fig:fig3.2}

Specific regions of each $B^\pm \to h^\pm h^{'+}h^{'-}$ decay mode are chosen to perform the studies. The rescattering region is defined in the Dalitz plot in the mass range 1.1-2.25 GeV$^2/c^2$ for the two-kaon for $B^\pm \to K^\pm K^+K^-$ and the mass region  1.0-2.25 GeV$^2/c^2$, for the other three channels. Most of the rescattering regions present a $CP$ asymmetry values in excess of at least five Gaussian standard deviations. High mass regions are also defined for $B^\pm \to \pi^\pm \pi^+\pi^-$ and $B^\pm \to \pi^\pm K^+K^-$. The total set of regions studied are summarized in Table~\ref{tab:tab3.1}. 

\begin{table}[hbt]
    \centering
    \caption{Definition of the phase-space regions for the $B^\pm \to h^\pm h^{'+}h^{'-}$ channels in units of GeV$^2/c^{4}$. The $\phi(1020)$ meson is excluded from $B^\pm \to K^\pm K^+K^-$.}
    \label{tab:tab3.1}
\begin{tabular}{c c c c}
\hline
& &  $ { B\xspace\xspace^\pm\xspace \rightarrow\xspace \pi\xspace\xspace^\pm\xspace \pi\xspace\xspace^+\xspace \pi\xspace\xspace^-\xspace}$ \xspace & \\
Region 1 & $\quad 1 < m^2(\pi^+ \pi^-)_{\textrm{low}} <$ 2.25 & and & $      3.5 < m^2(\pi^+ \pi^-)_{\textrm{high}} < 16$ \\
Region 2 & $\quad 1 < m^2(\pi^+ \pi^-)_{\textrm{low}} <$ 2.25 & and & $\;    16  < m^2(\pi^+ \pi^-)_{\textrm{high}} < 23$ \\
Region 3 & $4       < m^2(\pi^+ \pi^-)_{\textrm{low}} <$ 15   & and & $\quad 4   < m^2(\pi^+ \pi^-)_{\textrm{high}} < 16$ \\
\hline
& &  $ { B\xspace\xspace^\pm\xspace \rightarrow\xspace  K\xspace\xspace^\pm\xspace \pi\xspace\xspace^+\xspace \pi\xspace\xspace^-\xspace}$ \xspace & \\
Region 1 & $1 < m^2(\pi^+ \pi^-) < 2.25$ & and & $\; 3.5 < m^2(K^+ \pi^-) < 19.5$ \\
Region 2 & $1 < m^2(\pi^+ \pi^-) < 2.25$ & and & $19.5 < m^2(K^+ \pi^-) < 25.5$ \\
\hline
& &  $ { B\xspace\xspace^\pm\xspace \rightarrow\xspace \pi\xspace\xspace^\pm\xspace  K\xspace\xspace^+\xspace  K\xspace\xspace^-\xspace }$ \xspace & \\
Region 1 & $\quad 1 < m^2(K^+K^-) < 2.25$ & and & $4 < m^2(K^+ \pi^-) < 19$ \\
Region 2 & $4 < m^2(K^+K^-) < 25$         & and & $3 < m^2(K^+ \pi^-) < 16$  \\
\hline
& &  $ { B\xspace\xspace^\pm\xspace \rightarrow\xspace  K\xspace\xspace^\pm\xspace  K\xspace\xspace^+\xspace  K\xspace\xspace^-\xspace}$ \xspace & \\
Region 1 & $1.1 < m^2(K^+ K^-)_{\textrm{low}} < 2.25$  & and & $\; 4 < m^2(K^+ K^-)_{\textrm{high}} < 17$ \\
Region 2 & $1.1 < m^2(K^+ K^-)_{\textrm{low}} < 2.25$  & and & $17 < m^2(K^+ K^-)_{\textrm{high}} < 23$ \\
\hline
\end{tabular}
\end{table}

\begin{figure}[htb]
\centering  
\includegraphics[width=0.5
\linewidth]{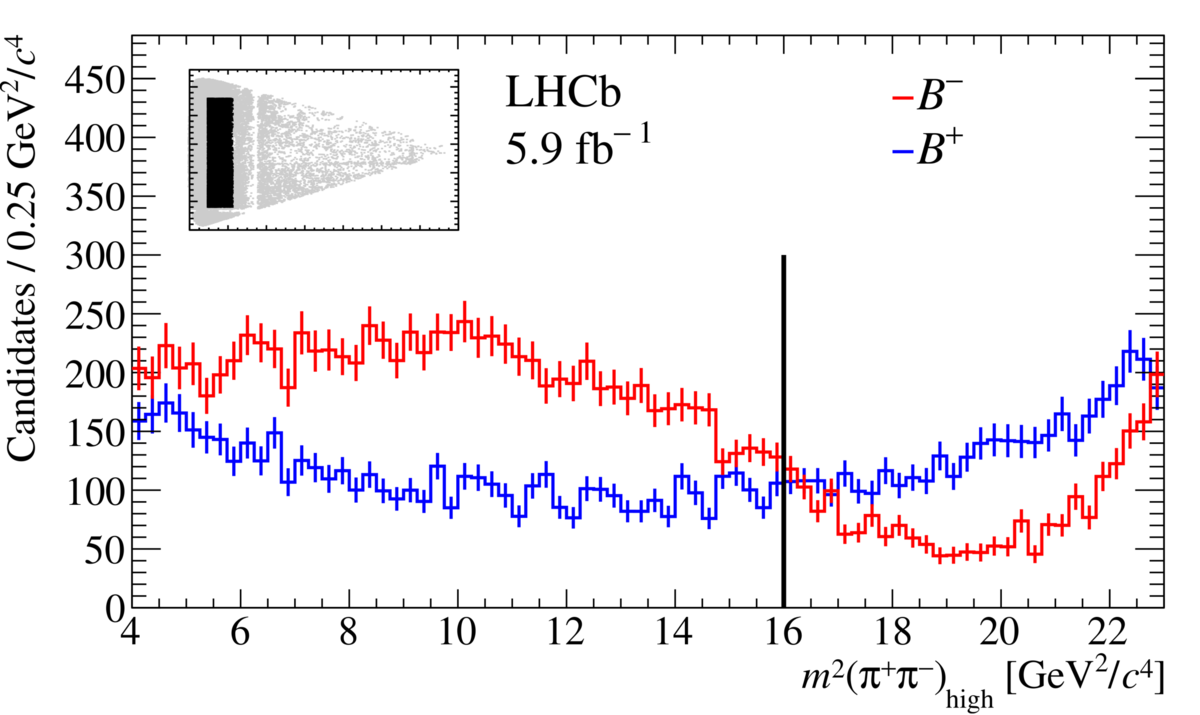}\\
\includegraphics[width=0.48
\linewidth]{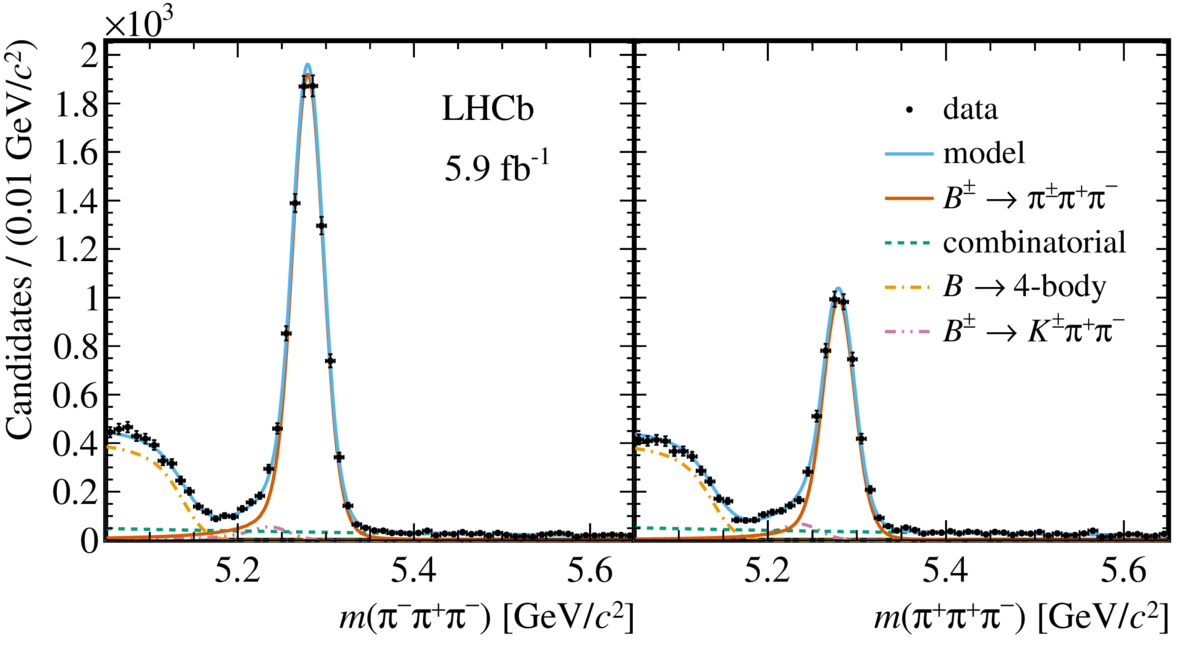}
\includegraphics[width=0.48
\linewidth]{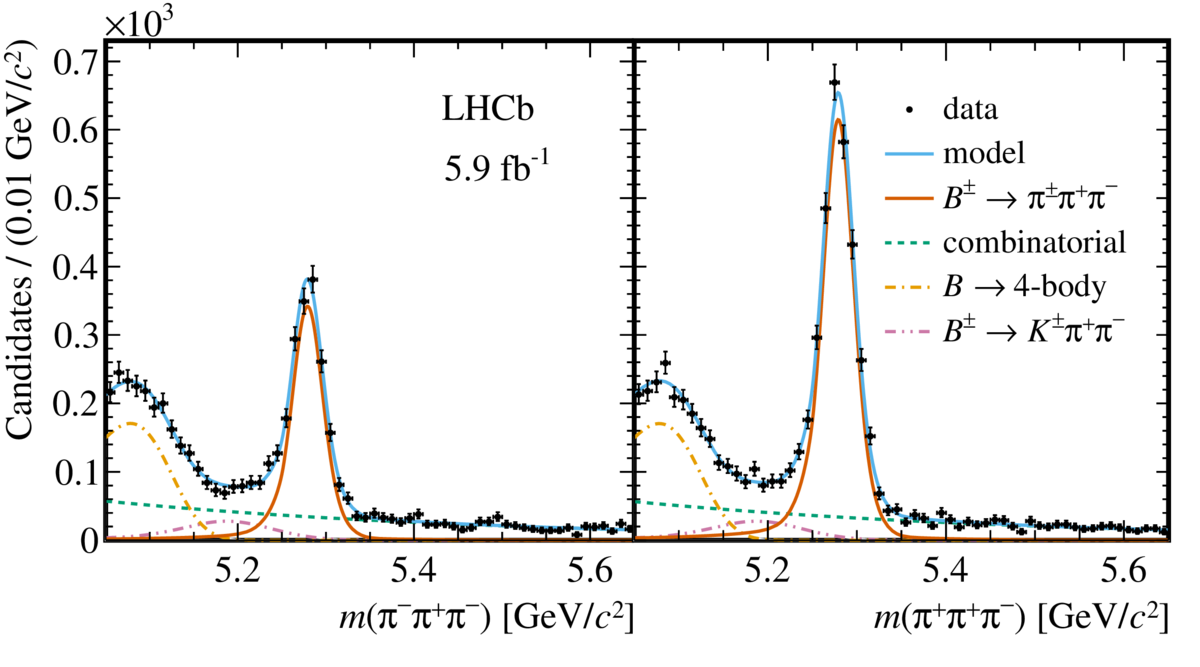}

\caption{(a) Projection of the $m^2(\pi^+\pi^-)_{high}$ rescattering region for (b) region 1 and (c) region 2. The regions are separated by the black vertical line in plot (a).}
\label{fig:fig3.3} 
\end{figure}

The $A_{CP}$ measurements in each region are shown in Table~\ref{tab:tab3.2}. As an example, the projections of the squared invariant mass $m^2(\pi^+\pi^-)_{high}$ for $B^\pm \to \pi^\pm \pi^+\pi^- $, the sub index high for being a symmetrical channel, in the rescattering region can be seen in Fig.~\ref{fig:fig3.3}. The projection shows that the $CP$ asymmetry is positive in region 1 and negative in region 2.


\begin{table}[hbt]
    \caption{Signal yield, raw asymmetry and $A_{CP}$ in the regions defined for each channel as give in Table~\ref{tab:tab3.1}}
    \label{tab:tab3.2}

    \centering
    {
\footnotesize

\begin{tabular}{c c c c}
\hline 
 $ { B\xspace\xspace^\pm\xspace \rightarrow\xspace \pi\xspace\xspace^\pm\xspace \pi\xspace\xspace^+\xspace \pi\xspace\xspace^-\xspace}$ \xspace & $N_{\text{sig}}$ & $A_{\text{raw}}$ &  $ A_{ C\!P\xspace}$ \xspace \\
Region 1 & $14\,330 \pm 150$ & $+$0.309 $\pm$ 0.009 & $+$0.303 $\pm$ 0.009 $\pm$ 0.004 $\pm$ 0.003 \\
Region 2 &  $\,\;4\,850 \pm 130$ & $-$0.287 $\pm$ 0.017 & $-$0.284 $\pm$ 0.017 $\pm$ 0.007 $\pm$ 0.003 \\
Region 3 &  $2\,270 \pm 60$  & $+$0.747 $\pm$ 0.027 & $+$0.745 $\pm$ 0.027 $\pm$ 0.018 $\pm$ 0.003 \\
\hline
 $ { B\xspace\xspace^\pm\xspace \rightarrow\xspace  K\xspace\xspace^\pm\xspace \pi\xspace\xspace^+\xspace \pi\xspace\xspace^-\xspace}$ \xspace & & & \\
Region 1 & $41\,980 \pm 280$  & $+$0.201 $\pm$ 0.005 & $+$0.217 $\pm$ 0.005 $\pm$ 0.005 $\pm$ 0.003 \\
Region 2 & $27\,040 \pm 250$  & $-$0.149 $\pm$ 0.007 & $-$0.145 $\pm$ 0.007 $\pm$ 0.006 $\pm$ 0.003 \\
\hline
 $ { B\xspace\xspace^\pm\xspace \rightarrow\xspace \pi\xspace\xspace^\pm\xspace  K\xspace\xspace^+\xspace  K\xspace\xspace^-\xspace }$ \xspace & & & \\
Region 1 & $11\,430 \pm 170$  & $-$0.363 $\pm$ 0.010 & $-$0.358 $\pm$ 0.010 $\pm$ 0.014 $\pm$ 0.003 \\
Region 2 & $\,\;2\,600  \pm 120$  & $+$0.075 $\pm$ 0.031 & $+$0.097 $\pm$ 0.031 $\pm$ 0.005 $\pm$ 0.003 \\
\hline
 $ { B\xspace\xspace^\pm\xspace \rightarrow\xspace  K\xspace\xspace^\pm\xspace  K\xspace\xspace^+\xspace  K\xspace\xspace^-\xspace}$ \xspace & & & \\
Region 1 & $76\,020 \pm 350$ & $-$0.189 $\pm$ 0.004 & $-$0.178 $\pm$ 0.004 $\pm$ 0.004 $\pm$ 0.003 \\
Region 2 & $37\,440 \pm 320$ & $+$0.030 $\pm$ 0.005 & $+$0.043 $\pm$ 0.005 $\pm$ 0.004 $\pm$ 0.003 \\
\hline
\end{tabular}
}
\end{table}

In the high mass region for $B^\pm \to \pi^\pm \pi^+\pi^- $, a large asymmetry is observed, where it can be seen the indication of $\chi_{c0}(1P)$ contribution around 11.6 \rm{GeV}$^2/c^4$. This is an interesting result as some reference have predicted $CP$ violation involving this decay~\cite{Gronau:1995hn, Bediaga:1998}. In general, it is observed large $CP$ asymmetries in localized regions of the phase space, with positive ad negative $CP$ asymmetries in the same $B$ charged channel. Results of the previous LHCb analysis are confirmed~\cite{LHCb:hhh2014}. It is also found large $CP$ violation in the $\pi\pi \leftrightarrow KK$ rescattering region.

\subsubsection*{Search for  direct CPV in $B \to PV$}
The LHCb collaboration also reports the $CP$ asymmetry measurements in charmless $B \to PV$ decays, where $P$ stands for pseudoscalar and $V$ for vector meson, using a method that does not required full amplitude analysis~\cite{Alvarenga:2016}. The dataset used is the full Run 2 data collected between 2015 and 2018 corresponding to an integrated luminosity of $5.9$ fb$^{-1}$.

The method is based in three key features of three-body $B$ decays: the large phase space, the dominance of scalar and vector resonances with masses below or around 1 GeV/$c^{2}$ and the clear signatures of resonances in the phase space~\cite{Bediaga:2009,Bigi:1509,Grossman:2021}. For a decay $B^\pm \to R(\to h_1^-h_2^+)h^\pm_3$, where R is the resonance, the following notation is used: 
\begin{itemize}
    \item $s_{||}$ for the two-body invariant mass squared $m^2(h_1^{-}h_2^{+})$
    \item $s_{\perp}$ for the two-body invariant mass squared $m^2(h_1^{-}h_3^{+})$
\end{itemize}
 
The decay amplitudes for $B^+$ and $B^-$ are usually represented as the coherent sum of intermediate states. For the case that one vector resonance is interfering with a scalar component, the decay amplitude can be expressed as
\begin{equation}
 \mathcal{M}_{\pm} = a_{\pm}^Ve^{i\delta_\pm^V}F_V^{BW} \rm{cos} \theta(s_{\perp}, s_{||})+a_{\pm}^Se^{i\delta_\pm^S}F_S^{BW},   
\end{equation}
where $a_{\pm}^V$ and $a_{\pm}^S$ are the magnitudes of the vector and scalar resonances, respectively, assumed to be independent of $s_{\perp}$. $\delta_\pm^V$ and $\delta_\pm^S $ are the phases of the vector and scalar amplitudes, and $\theta(s_{\perp}, s_{||})$ is the helicity angle. The resonances may be described by a Breit-Wigner (BW) function, $F_R^{BW}$. Assuming that $a_{\pm}^V$, $a_{\pm}^S$ and $\delta_\pm^V$ and $\delta_\pm^S$ do not depend on $s_{\perp}$, the squared decay amplitude can be expressed as quadratic polynomial in \rm{cos}$\theta(m^2_V, s_{\perp})$ as

\begin{equation}
    |\mathcal{M}_{\pm}|^2 = f(\rm{cos}\theta(m^2_V, s_{\perp})) = p_0^{\pm} + p_1^\pm \rm{cos}\theta(m^2_V, s_{\perp}) + p_2^\pm \rm{cos}^2\theta(m^2_V, s_{\perp}),
    \label{eq:eq1}
\end{equation}
where $p_0^\pm$, $p_1^\pm$, $p_2^\pm$ are polynomial coefficients. The $CP$ asymmetry $A_{CP}^V$ in the $B \to PV$ decay is given then as function of $p_2^\pm$,

\begin{equation}
    A_{CP}^V = \frac{|\mathcal{M}_-|^2 - |\mathcal{M}_+|^2}{|\mathcal{M}_-|^2 + |\mathcal{M}_+|^2} = \frac{p_2^- - p_2^+}{p_2^- + p_2^+}
\end{equation}

The histograms of data projected in the $s_{\perp}$ axes is fitted  with the function defined in Eq.~\eqref{eq:eq1}. Then, the parameters necessary to compute the $A_{CP}$ are obtained.
The regions studied for each  $B^\pm \to h^\pm h^{'+}h^{'-}$ mode are the following:
\begin{itemize}
    \item For $B^\pm \to \pi^\pm \pi^+\pi^-$, the region $B^\pm \to \rho(770)^0\pi^\pm$.

    \item For $B^\pm \to K^\pm \pi^+\pi^-$ the regions $B^\pm \to \rho(770)^0\pi^\pm$ and  $B^\pm \to \overline{K}^{*}(892)^0\pi^\pm$. 

    \item For $B^\pm \to K^\pm K^+K^-$ the region $B^\pm \to \phi(1020)K^\pm$. 

    \item For $B^\pm \to \pi^\pm K^+K^-$ the region $B^\pm \to \overline{K}^{*}(892)^0\pi^\pm$. 
\end{itemize}

\begin{figure}[htb]
\centering  
\includegraphics[width=0.42
\linewidth]{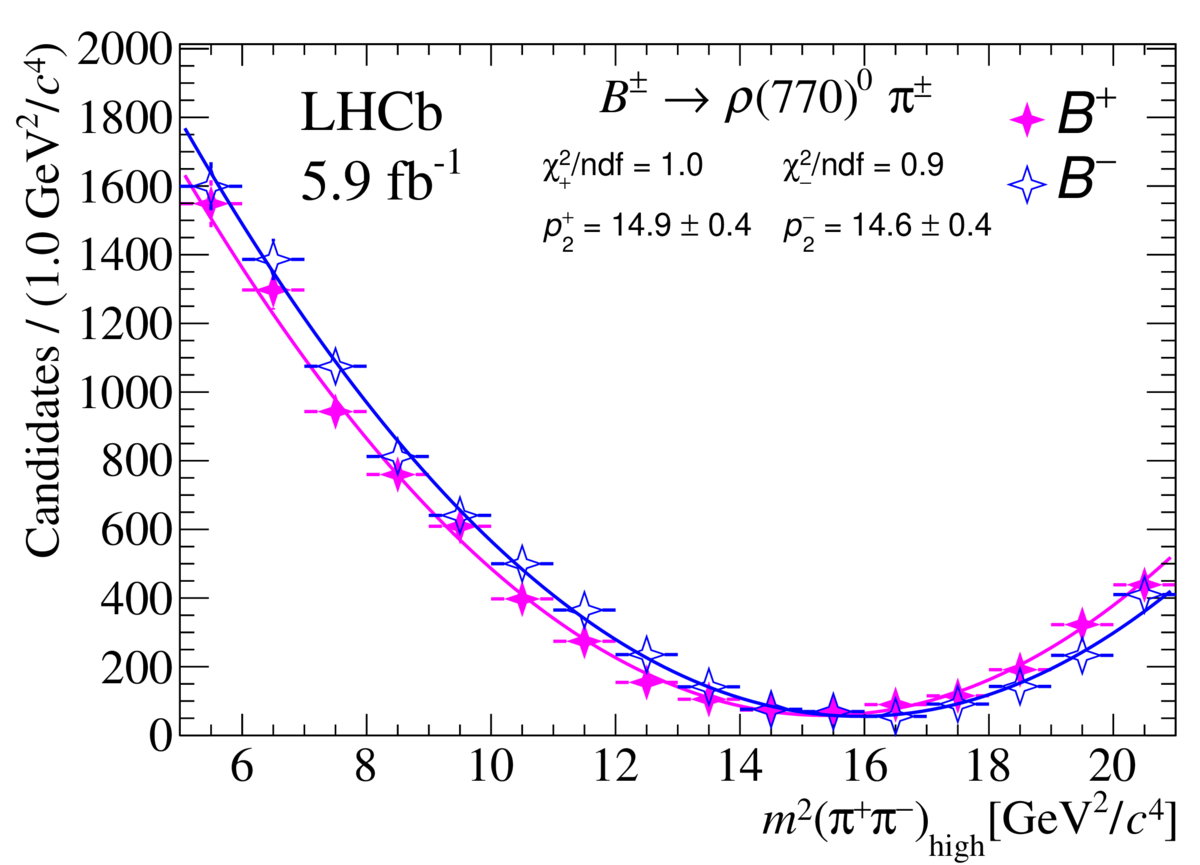}
\includegraphics[width=0.42
\linewidth]{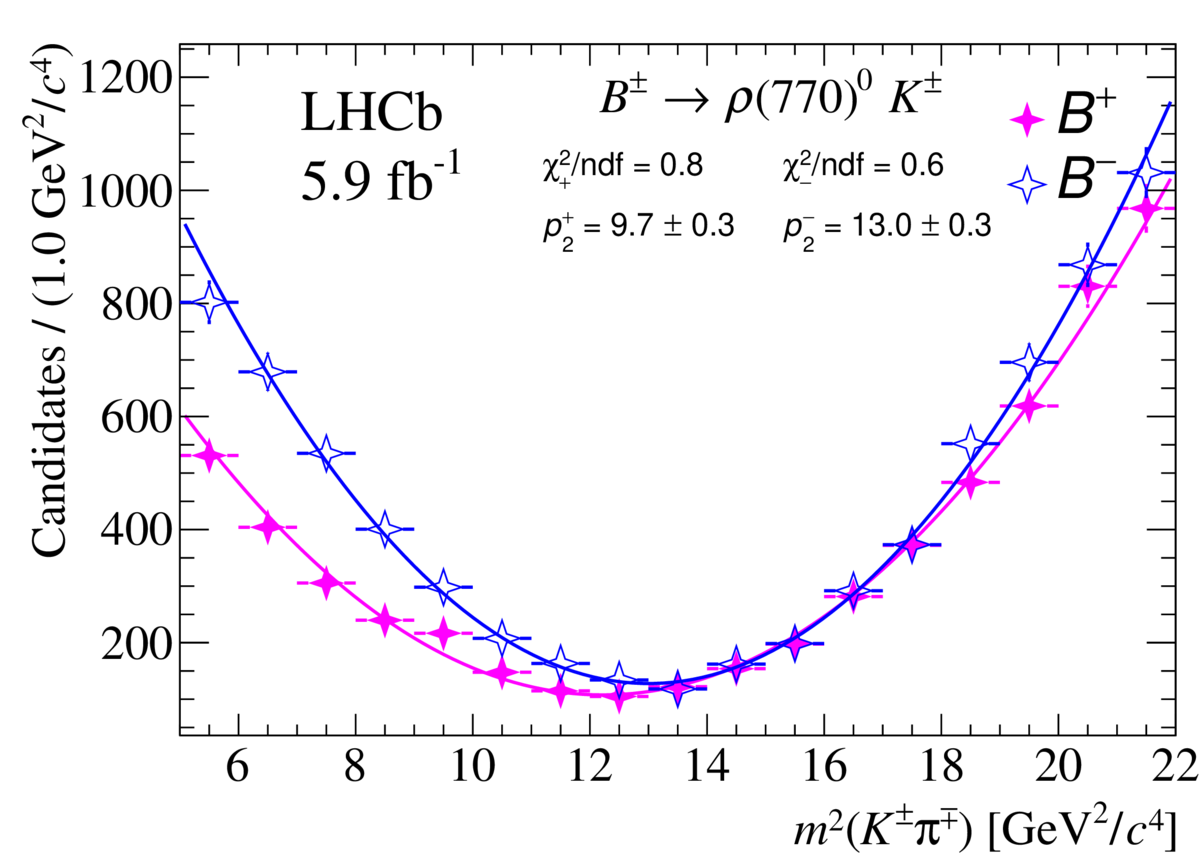}
\includegraphics[width=0.42
\linewidth]{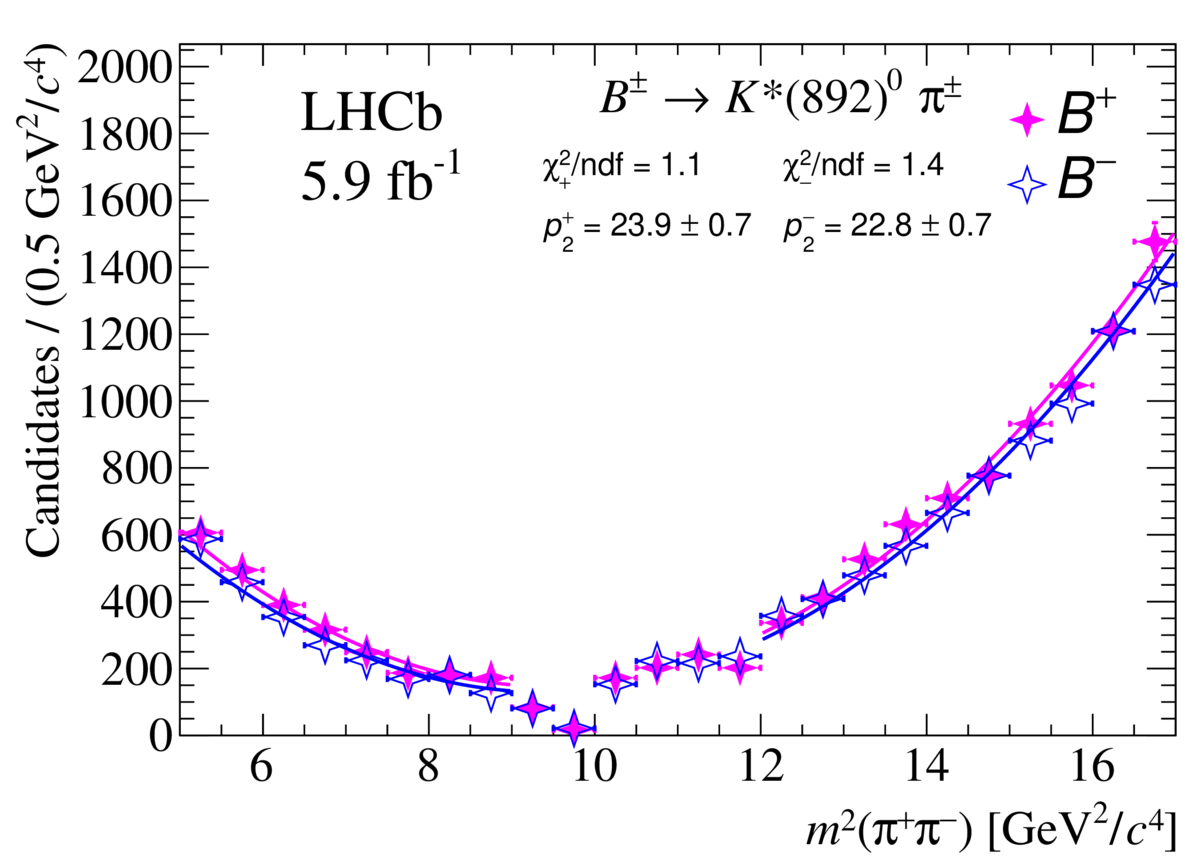}
\includegraphics[width=0.42
\linewidth]{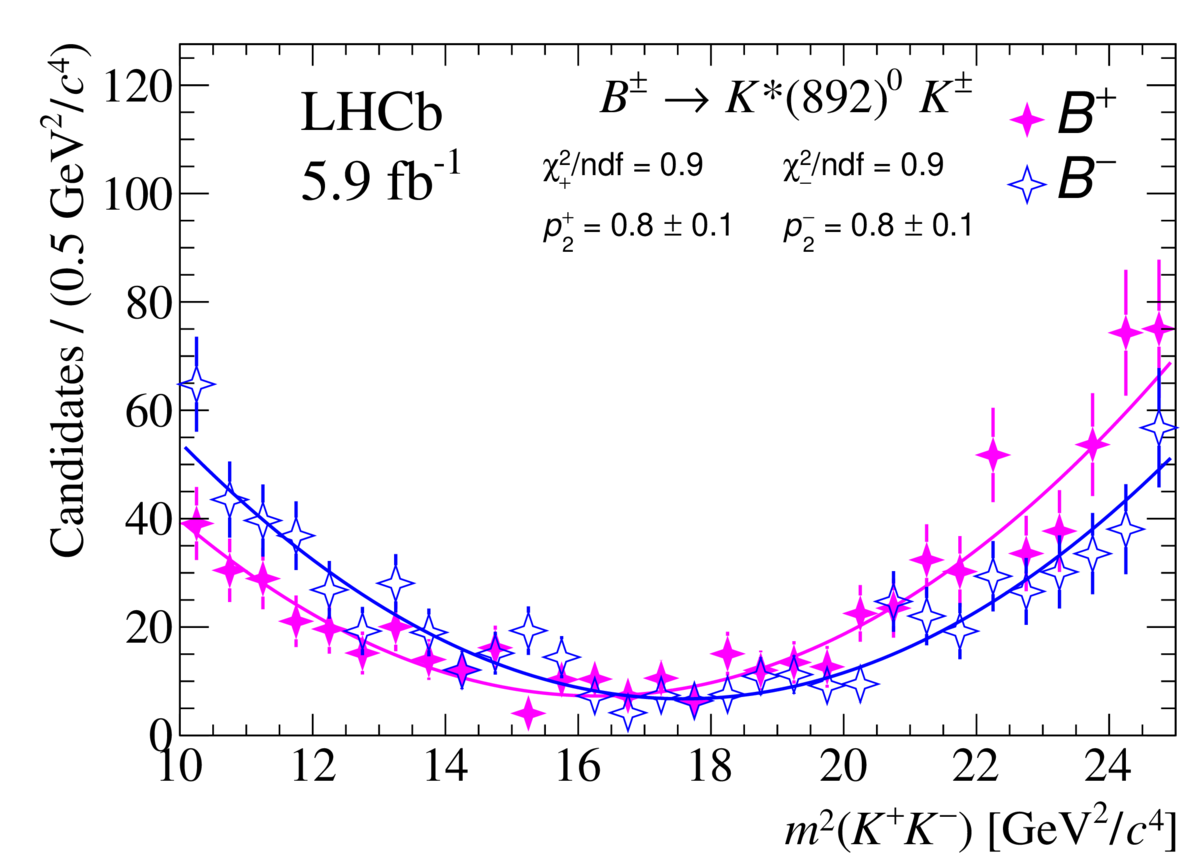}
\includegraphics[width=0.42
\linewidth]{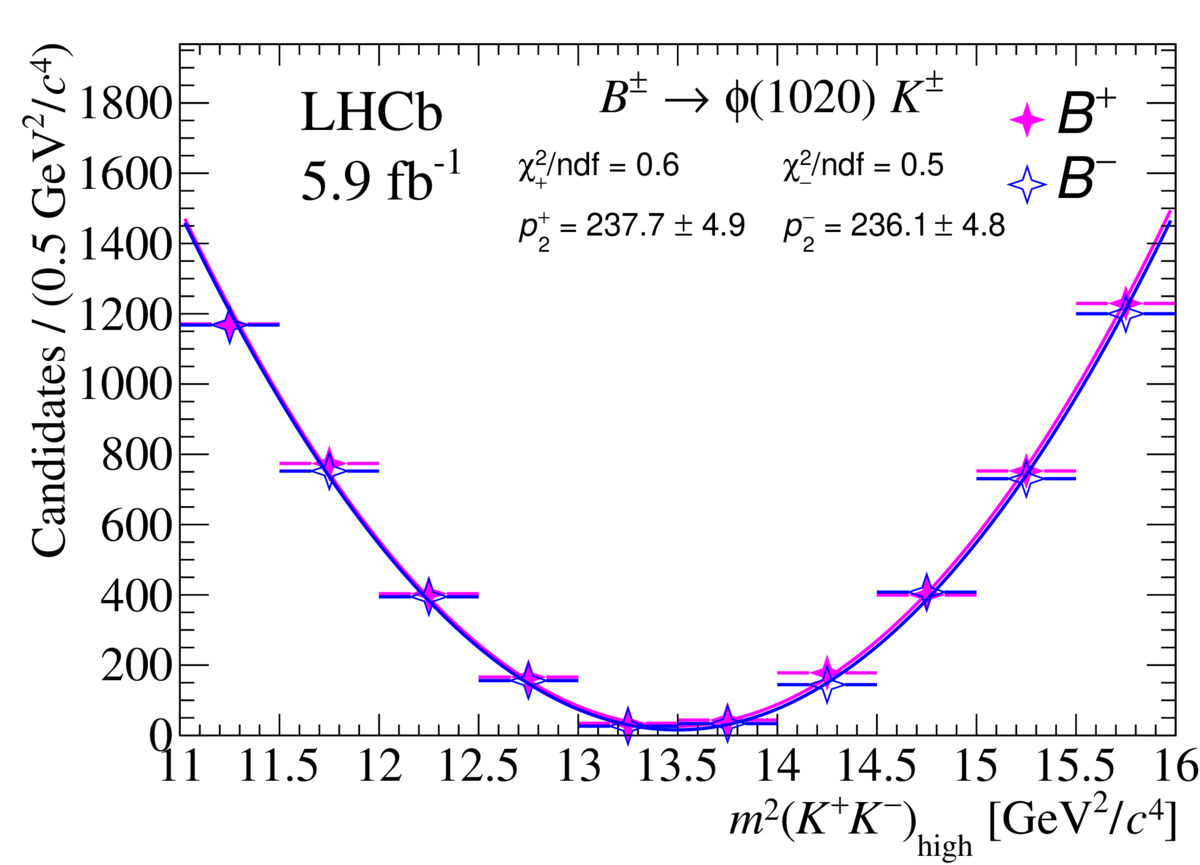}
\caption{Projection of the fit model to the data distribution for the regions selected for each channel. (a) $\rho(770)^0$ in $B^\pm \to \pi^\pm \pi^+\pi^-$, (b) $\rho(770)^0$ in $B^\pm \to K^\pm \pi^+\pi^-$, (c) $\overline{K}^*(892)^0$ in $B^\pm \to K^\pm \pi^+\pi^-$, (d) $\overline{K}^*(892)^0$ in $B^\pm \to \pi^\pm K^+K^-$ and (e) $\phi(1020)$ in $B^\pm \to K^\pm K^+K^-$ ~\cite{LHCb:BPV}.\label{fig:fig3.4}} 
\end{figure}

The fits to the distribution of $s_{\perp}$ for $B^+$ and $B^-$ can be seen in Fig.~\ref{fig:fig3.4} for the regions just described above.

The summary of the $A_{CP}$ measurements for the five decay channel is presented in Table~\ref{tab:tab3.4}~\cite{LHCb:BPV}. For all channels, the $CP$ asymmetries are compatible with zero, except the $CP$ asymmetry for $B^\pm \to \rho(770)^0 K^\pm$ region, $A_{CP} = +0.150 \pm 0.019 \pm 0.011$, which differs from zero with a significance of 6.8$\sigma$. This is the first observation of $CP$ asymmetry in this decay mode. 

\begin{table}[tbh]
    \caption{ $CP$-asymmetry measurements for the vector resonance channels in their associated final state $B^\pm \to R(\to h_{1}^{-}h_{2}^{+})h_3^{\pm}$ decays. Previous measurements from others experiments are also included~\cite{LHCb:BPV}.\label{tab:tab3.4}}
    
    \centering
\footnotesize

\begin{tabular}{lcl}
\toprule
Decay channel  & This work & Previous measurements \\ 
\midrule
$B^\pm \rightarrow\xspace$( $\rho(770)^0$ \xspace $\rightarrow\xspace \pi^+ \pi^-$)$\pi^\pm$        & $-$0.004 $\pm$ 0.017 $\pm$ 0.009 & $+$0.007 $\pm$ 0.011 $\pm$ 0.016   \\ \midrule
$B^\pm \rightarrow\xspace$( $\rho(770)^0$ \xspace $\rightarrow\xspace \pi^+ \pi^-$)$K^\pm$        & $+$0.150 $\pm$ 0.019 $\pm$ 0.011 & $+$0.44  $\pm$ 0.10  $\pm$ 0.04   \\
		                                &                                  & $+$0.30  $\pm$ 0.11  $\pm$ 0.02   \\ \midrule
$B^\pm \rightarrow\xspace$( $\kern 0.18em\optbar{\kern -0.18em  K\xspace}{}\xspace^*(892)^0$ \xspace $\rightarrow\xspace K^{\pm} \pi^{\mp}$)$\pi^\pm$ & $-$0.015 $\pm$ 0.021 $\pm$ 0.012 & $+$0.032 $\pm$ 0.052 $\pm$ 0.011  \\ 
		                                &                                  & $-$0.149 $\pm$ 0.064 $\pm$ 0.020  \\ \midrule
$B^\pm \rightarrow\xspace$( $\kern 0.18em\optbar{\kern -0.18em  K\xspace}{}\xspace^*(892)^0$ \xspace $\rightarrow\xspace K^{\pm} \pi^{\mp}$)$K^\pm$ & $+$0.007 $\pm$ 0.054 $\pm$ 0.032 & $+$0.123 $\pm$ 0.087 $\pm$ 0.045   \\ \midrule
$B^\pm \rightarrow\xspace$( $\phi(1020)$ \xspace  $\rightarrow\xspace K^+ K^-$)$K^\pm$           & $+$0.004 $\pm$ 0.014 $\pm$ 0.007 & $+$0.128 $\pm$ 0.044 $\pm$ 0.013  \\ 
\bottomrule
\end{tabular}
\end{table}

\subsection{Search for $CP$ Violation in Baryons }

The violation of the Charge-Parity asymmetry has been observed in $K$ and $B$ mesons decays~\cite{Christenson:1964fg, BaBar:2001ags, Belle:2001zzw} and was recently also observed in $D$ mesons by the LHCb collaboration~\cite{LHCb:2019hro}. In the baryonic sector, although predicted, observation is yet to be claimed. The first
evidence was found by the LHCb experiment in the decay mode $\Lambda_b^0 \to p \pi^-\pi^+\pi^-$~\cite{LHCb:2016yco} with a significance of 3.3 standard deviation including systematic uncertainties. Given the large production cross-section of beauty baryons in $pp$ collisions at the Large Hadron Collider (LHC), promising studies can be conducted within the scope of the LHCb experiment. Several new analysis are reported in the following.

With a data collected between 2011 to 2017, corresponding to an integrated luminosity of 6.6 fb$^{-1}$, the decay channel $\Lambda_b^0 \to p \pi^-\pi^+\pi^-$ is studied to search for $CP$ violation and observation of $P$ violation, which supersede previous results~\cite{LHCb:2016yco}. This decay mode is mainly mediated by tree and loop processes of similar magnitude, allowing for significant interference effects. It posses a rich resonant structure, with the dominant contribution proceeding through $\Lambda_b^0 \to N^{*+}\pi^-$, with $N^{*+} \to \Lambda^{++}(1234) \pi^-$ and  $\Lambda^{++}(1234) \to p\pi^+$, or as well $\Lambda_b^0 \to p a_1^{-}(1260)$, with $a_1^{-}(1260) \to \rho^0(770)\pi^-$ and $\rho^0(770) \to \pi^+\pi^-$ decays, where the proton excited states are indicated as $N^{*+}$.

The analysis have been performed following two strategies: by using the triple product asymmetries (TPA) in order to search for $CP$ and $P$ violation, and applying the unbinned energy test method. The searches for $CP$ violation are performed by separating the $P$-odd and $P$-even contributions. In the TPA analysis, both local and integrated asymmetries are considered. The scalar triple product, built in the $\Lambda_b^0$ rest frame, is defined as
\begin{align}
    C_{\hat{T}} \equiv \vec{p}_p \cdot (\vec{p}_{\pi^-_{\rm{fast}} \times \vec{p}_{\pi^+}}),   \quad\overline{C}_{\hat{T}} \equiv \vec{p}_{\bar{p}} \cdot (\vec{p}_{\pi^-_{\rm{fast}}} \times \vec{p}_{\pi^-})
\end{align}
for $\Lambda_b^0$ and $\bar{\Lambda}_b^0$ respectively. The $\pi_{fast}^-$ ($\pi_{slow}^{-}$) refers to faster or slower of the two negative pions in the $\Lambda_b^0$ rest frame. In total four statistically independent subsamples are considered, labeled $I$ for $C_{\hat{T}} >0$, $II$ for $C_{\hat{T}} <0$, $III$ for $-\bar{C}_{\hat{T}} >0$ and $IV$ for $-\bar{C}_{\hat{T}} <0$. Both $CP$- and $P$ violating effects appears as differences between the triple product observables related by $CP$ and $P$ transformations. The TPA are defined as

\begin{align}
    A_{\hat{T}} = \frac{N(C_{\hat{T}} > 0) - N(C_{\hat{T}} < 0)}{N(C_{\hat{T}} > 0) + N(C_{\hat{T}} < 0)},\quad \overline{A}_{\hat{T}} = \frac{\bar{N}(-\bar{C}_{\hat{T}} > 0) - \bar{N}(-\bar{C}_{\hat{T}} < 0)}{\bar{N}(-\bar{C}_{\hat{T}} > 0) + \bar{N}(-\bar{C}_{\hat{T}} < 0)}, 
\end{align}
 where $N$ and $\bar{N}$ are the signal yields for $\Lambda_b^0$ and $\bar{\Lambda}_b^0$ respectively. Finally, the $CP$- and $P$-violating asymmetries are defined as

 \begin{align}
     a_{CP}^{\hat{T}-odd} = \frac{1}{2}(A_{\hat{T}} - \overline{A}_{\hat{T}}), \quad\quad a_{P}^{\hat{T}-odd} = \frac{1}{2}(A_{\hat{T}} + \overline{A}_{\hat{T}})
 \end{align}
 
 Two binning schemes are used, the first one (A) is motivated on the results of an approximate amplitude analysis of $\Lambda_b^0 \to p \pi^-\pi^+\pi^-$ decays. A second binning scheme (B) is used to probe the asymmetries as function of $|\phi|$,  the absolute value of the angle between the planes defined by $p \pi_{fast}^-$ and $\pi^+\pi^-_{slow}$ systems. It is also taken into account the invariant-mass regions $m(p\pi^+\pi^-_{slow}) > 2.8$ GeV$/c^2$ and $m(p\pi^+\pi^-_{slow}) < 2.8$ GeV$/c^2$, where the $a_1$ resonance and $N^{*+}$ decay dominate, respectively. The total samples studied are labeled then, respectively, A$_1$, B$_1$, A$_2$ and B$_2$.

 The energy test is a model-independent unbinned test, which is sensitive to local differences between two samples. e.g as could arise from $CP$ violation. It is performed by the calculation of a statistic test:
\begin{equation}
    T \equiv \frac{1}{2n(n-1)}\sum_{i \neq j}^{n} \psi_{ij} + \frac{1}{2\bar{n}(\bar{n}-1)}\sum_{i\neq j}^n\psi_{ij} - \frac{1}{n\bar{n}}\sum_{i=1}^n\sum_{j=1}^{\bar{n}} \psi_{ij},
\end{equation}
where $n$ and $\bar{n}$ are the candidates in the first and second sample. Each pair of candidates $ij$ is assigned a weight $\psi_{ij} = e^{-d^2_{ij}/2\delta^2}$, where $d_{ij}$ is their Euclidean distance in phase space, $\delta$ determines the distance scale probed using the energy test.

The results of the measured TPA from the fit to the full data sample are~\cite{LHCb:2019oke}
\begin{center}
    $a_{CP}^{\hat{T}-odd} = (-0.7 \pm 0.7 \pm 0.2)$\% and $a_{P}^{\hat{T}-odd} = (-4.0 \pm 0.7 \pm 0.2)$\%
\end{center}
where the first result is consistent with the $CP$-conserving hypothesis. A nonzero value for $a_{P}^{\hat{T}-odd}$ is found, and with a significance of 5.5 standard deviations indicates parity violation in the $\Lambda_b^0 \to p \pi^-\pi^+\pi^-$ decay. 

\begin{figure}[htb]
\centering  
\includegraphics[width=0.45
\linewidth]{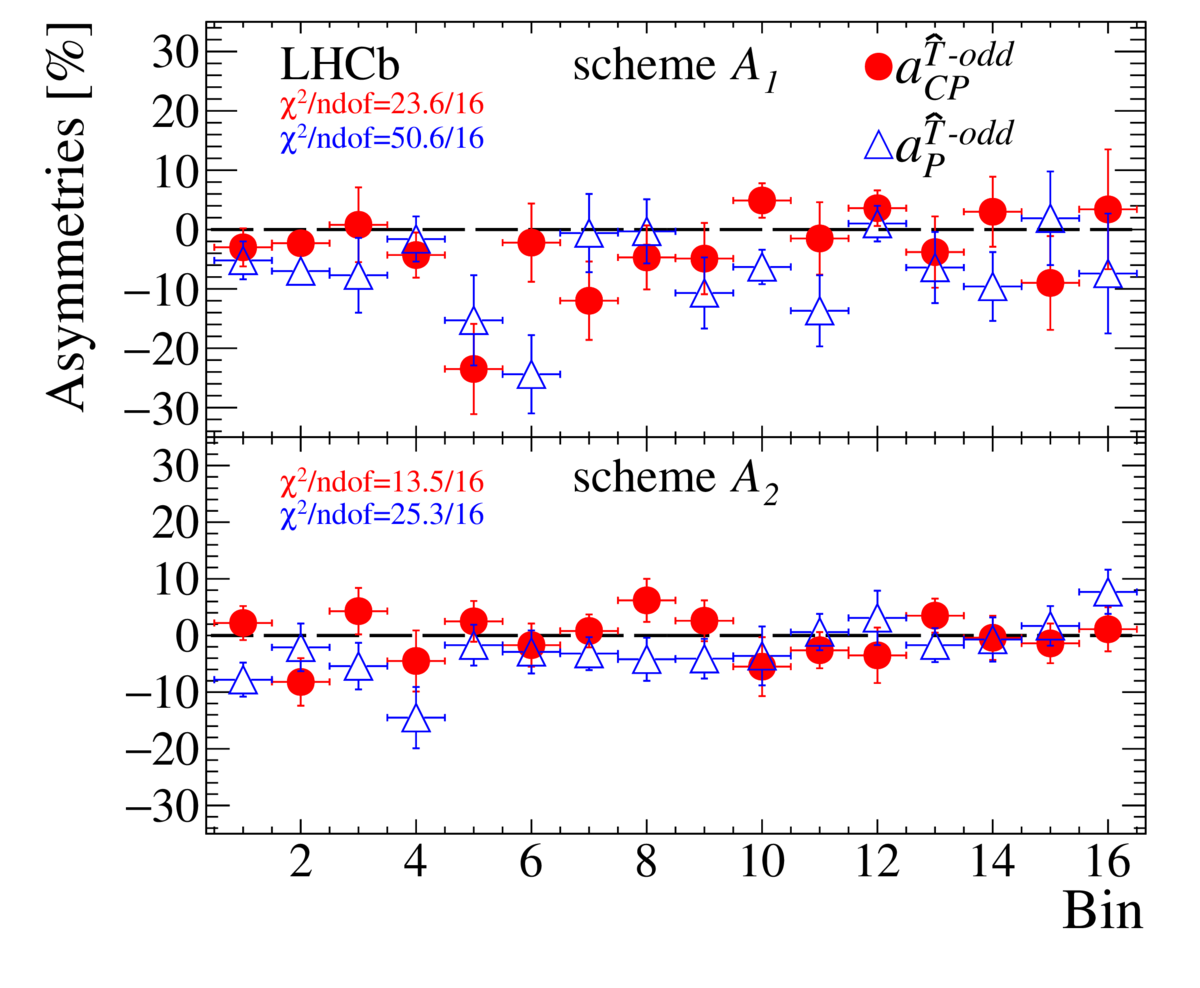}
\includegraphics[width=0.45
\linewidth]{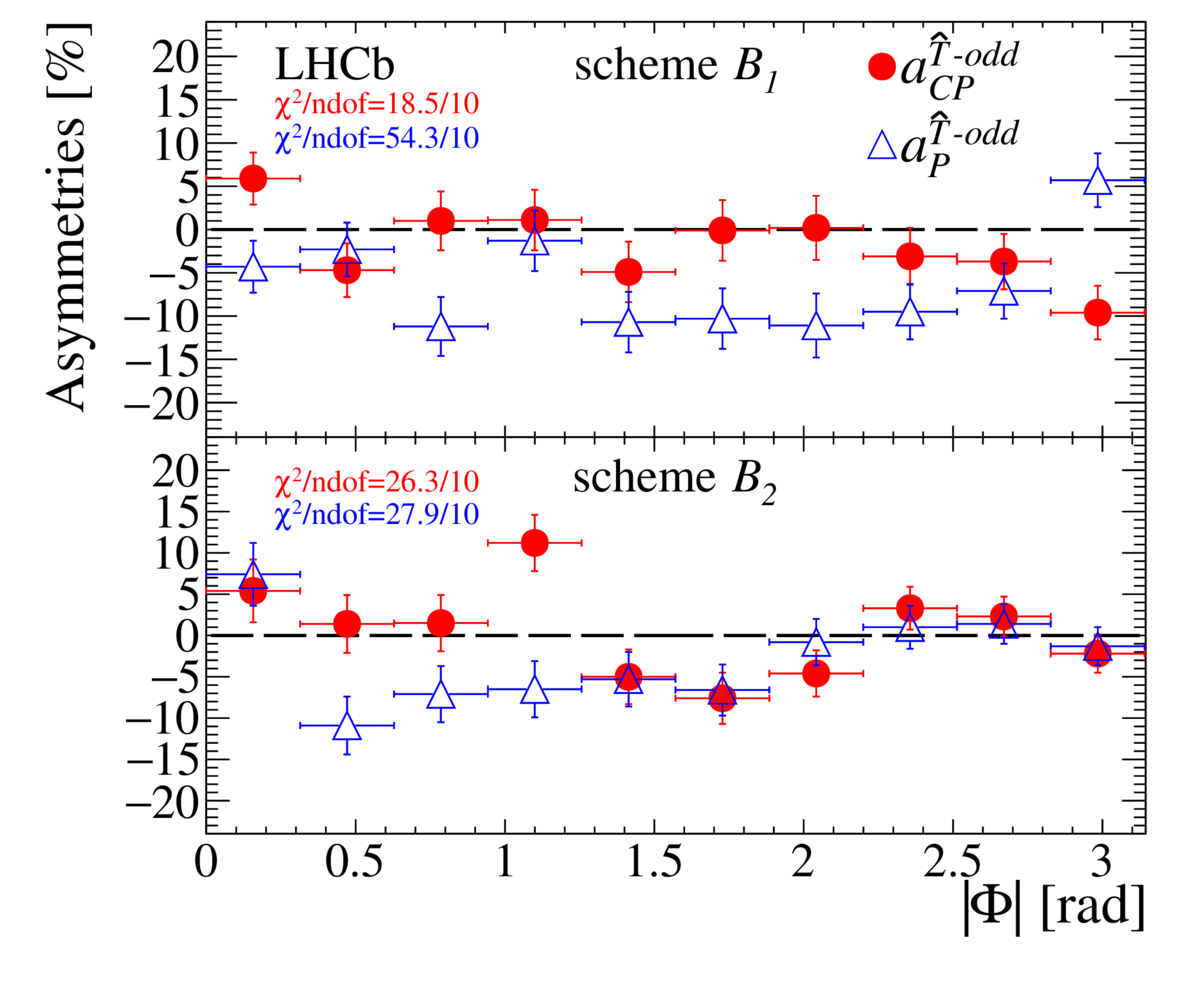}
\caption{Asymmetries results for the binning scheme A$_1$, A$_2$ (left) and B$_1$, B$_2$ (right) ~\cite{LHCb:2019oke}.\label{fig:fig3.4.1}}
\end{figure}

The values of the TPA for the binning schemes are shown in Fig.~\ref{fig:fig3.4.1}. The $A_2$ and $B_2$ phase space regions, the $p$-values with respect to the $CP$ conserving hypothesis corresponding to statistical significance of 0.5 and 2.9 standard deviations are measured, respectively. Therefore, the evidence of $CP$ violation previously observed~\cite{LHCb:2016yco} is not established.

The $p$-values obtained for the different configurations of the energy test can be seen in Table~\ref{tab:tab3.4.1}. All $CP$-violation searches result in $p$-values with a significance of 3 standard deviation or smaller. For both methods, the results are marginally compatible with the no $CP$-violation hypothesis. Parity violation is observed in both methods, locally with a significance over 5$\sigma$.

\begin{table}[hbt]
    \centering
\caption{Energy test results: $p$-values for different distance scales and test configurations~\cite{LHCb:2019oke}.}
 \label{tab:tab3.4.1}

\begin{tabular}{l|ccc}
Distance scale $\delta$       & 1.6 GeV$^2/c^4$  & 2.7 GeV$^2/c^4$ & 13 GeV$^2/c^4$   \\\hline
$p$-value ( $ C\!P$ \xspace conservation, $P$ even) & $3.1 \times 10^{-2}$ & $2.7 \times 10^{-3}$ & $1.3 \times 10^{-2}$\\
$p$-value ( $ C\!P$ \xspace conservation, $P$ odd) & $1.5 \times 10^{-1}$ & $6.9 \times 10^{-2}$ & $6.5 \times 10^{-2}$   \\
$p$-value ($P$ conservation) & $1.3 \times 10^{-7}$ & $4.0 \times 10^{-7}$ & $1.6 \times 10^{-1}$   \\
\end{tabular}
\end{table}

It was also reported by the LHCb collaboration the study of the $b$ baryonic decay $\Lambda_b^0 \to D^0pK^-$~\cite{LHCb:2021ohr}, to search for the decay  $D^0 \to K^+\pi^-$ and perform $CP$ violation measurements, where $D$ represents a $D^0$ or $\bar{D}^0$ meson. This mode is expected to be suppressed relative to the $\Lambda_b^0 \to [K^-\pi^+]_DpK^-$ decay. The processes $\Lambda_b^0 \to [K^-\pi^+]_DpK^-$ and $\Lambda_b^0 \to [K^+\pi^-]_DpK^-$ with same and opposite sign kaons are referred as the flavoured and suppressed decays, respectively. The Cabibbo-suppressed decay is particularly interesting as it has $b\to c$ and $b \to u$ contributing amplitudes in the same order of magnitude. The interference of these two amplitudes is sensitive to the CKM angle $\gamma$.

The analysis was performed with the full data sample collected by the LHCb detector from 2011-2018. The branching ratio of the decay mode $\Lambda_b^0 \to [K^-\pi^+]_DpK^-$ relative to $\Lambda_b^0 \to [K^+\pi^-]_DpK^-$ is measured, defined as
\begin{equation}
    R = \frac{\mathcal{B}(\Lambda_b^0 \to [K^-\pi^+]_DpK^-)}{\mathcal{B}(\Lambda_b^0 \to [K^+\pi^-]_DpK^-)},
\end{equation}
including both flavours. The $CP$ asymmetry in the suppressed mode is expressed as 

\begin{equation}
    A = \frac{\mathcal{B}(\Lambda_b^0 \to [K^+\pi^-]_DpK^-) - \mathcal{B}(\bar{\Lambda}_b^0 \to [K^-\pi^+]_DpK^+)}{\mathcal{B}(\Lambda_b^0 \to [K^+\pi^-]_DpK^-) + \mathcal{B}(\bar{\Lambda}_b^0 \to [K^-\pi^+]_DpK^+}
\end{equation}
The ratio of the branching fractions and the $CP$ asymmetry in the suppressed mode are measured separately in the phase space as well as in localized regions involving $\Lambda_b^0 \to DX$ decays, where an enhance sensitivity to $\gamma$ is expected.

The suppressed decay channel, $\Lambda_b^0 \to [K^+\pi^-]_DpK^-$, is observed for the first time with signal yield of $241 \pm 22$ events in full phase space. The branching ratio and $CP$ asymmetry are measured to be

\begin{equation*}
    R = 7.1 \pm 0.8 (\rm{stat.)^{+0.4}_{-0.3} (\rm{syst.})}, \quad
    A = 0.12 \pm 0.09 (\rm{stat.})^{+0.02}_{-0.03} (syst.).
\end{equation*}
In the restricted phase-space region, $M^2(pK^-) < 5$ GeV$^2/c^4$, the ratio and $CP$ asymmetry are measured to be

\begin{equation*}
    R = 8.6 \pm 1.5 (\rm{stat.)^{+0.4}_{-0.3} (\rm{syst.})}, \quad
    A = 0.01 \pm 0.16 (\rm{stat.})^{+0.03}_{-0.02} (syst.).
\end{equation*}
The ratio of the branching fractions is consistent with the estimated value from the relevant CKM matrix elements. The asymmetry values are consistent with zero in the full and restricted regions of the phase-space. The signal yields extracted with current available data sample is still too low to perform the measurement of the angle $\gamma$, but with the projected data sample to be collected by the LHCb in the upgrade I and later in the upgrade II, the study of this decay mode will contribute to the overall determination of $\gamma$.

Efforts have also been directed to search for $CP$ violation in the charmless decay $\Xi_b^{-} \to pK^-K^-$~\cite{LHCb:2021enr}. Tests have been performed on $\Lambda_b^0$ baryon decays to $p\pi^-$, $pK^-$, $K_s^0p\pi^-$, $\Lambda K^+K^-$, $\Lambda K^+\pi^-$, $p\pi^-\pi^+\pi^-$, $p\pi^-K^+K^-$, $pK^-\pi^+\pi^-$ and $pK^-K^+K^-$ final states.  The $\Xi_b^0$ decays has been studied in the processes $pK^-\pi^+\pi^-$ and $pK^-\pi^+K^-$. No $CP$ violation have been confirmed yet. Given the large $CP$ violation effects in charmless three-body $B$ decays, the study of the recently observed decay $\Xi_b^{-} \to pK^-K^-$, represents an interesting place to search for $CP$ violation effects.

The first amplitude analysis of a $b$ baryon decay, allowing for $CP$ violation effects was performed for $\Xi_b^{-} \to pK^-K^-$ using the Run 1, 2011-2012, and Run 2 data sample, 2015-2016, by the LHCb collaboration. A good description of data was obtained considering the resonant contributions of $\Sigma(1385)$, $\Lambda(1405)$, $\Lambda(1520)$, $\Lambda(1670)$, $\Sigma(1775)$ and $\Sigma(1915)$. The $CP$ asymmetry is evaluated for each contribution; no significant $CP$- violation effects are found. The decay modes $\Xi_b^- \to \Lambda(1520)K^-$ and $\Xi_b^- \to \Lambda(1670)K^-$ are observed with a significance greater than $5\sigma$. Branching fraction measurements, the establishment of the ratio of fragmentation and branching fractions of $\Omega_b^- \to p K^-K^-$ and $\Xi_b^- \to p K^-K^-$ have been performed~\cite{LHCb:2021enr}. With the upgrade of the LHCb, the study of three-body $b$ -baryon decays will be possible with larger samples projecting a overall rich program on the baryonic sector.

\section{Measurement of Branching ratios in hadronic $B$ to charm decays}
Hadronic $B$ decays accounting for ~75\% of the total branching fraction dominated by $b \to c$ processes, provides an opportunity to probe the SM by over-constraining the CKM triangle and via isospin sum rules. Effective hadronic $B$ tagging built on $B$ to charm decays allows neutrino reconstruction at Belle II, and plays a very crucial role for a large part of Belle II physics program. However, knowledge of $B$ meson hadronic decays is limited: about 40\% of the total $B$ width is not measured in terms of exclusive branching fractions ($B$), and thus is generated by the simulation with the Pythia fragmentation model~\cite{Bierlich:2022pfr}, which is known to be inaccurate.  Even among the measurements, most are performed with small data sets, hence have a large statistical uncertainties. So Poor knowledge of hadronic $B$ decays limits our reach to exciting physics. Belle II is (re)measuring many modes with the intention of improving MC (understanding). An example of remeasurement is $B \to D^{(*)}K K^{(*)}$ decays. The high purity of these decays makes them ideal candidates to improve the $B$-tagging efficiency. 

\begin{figure}[htb]
\centering
\subfigure{\includegraphics[width=0.49\columnwidth]{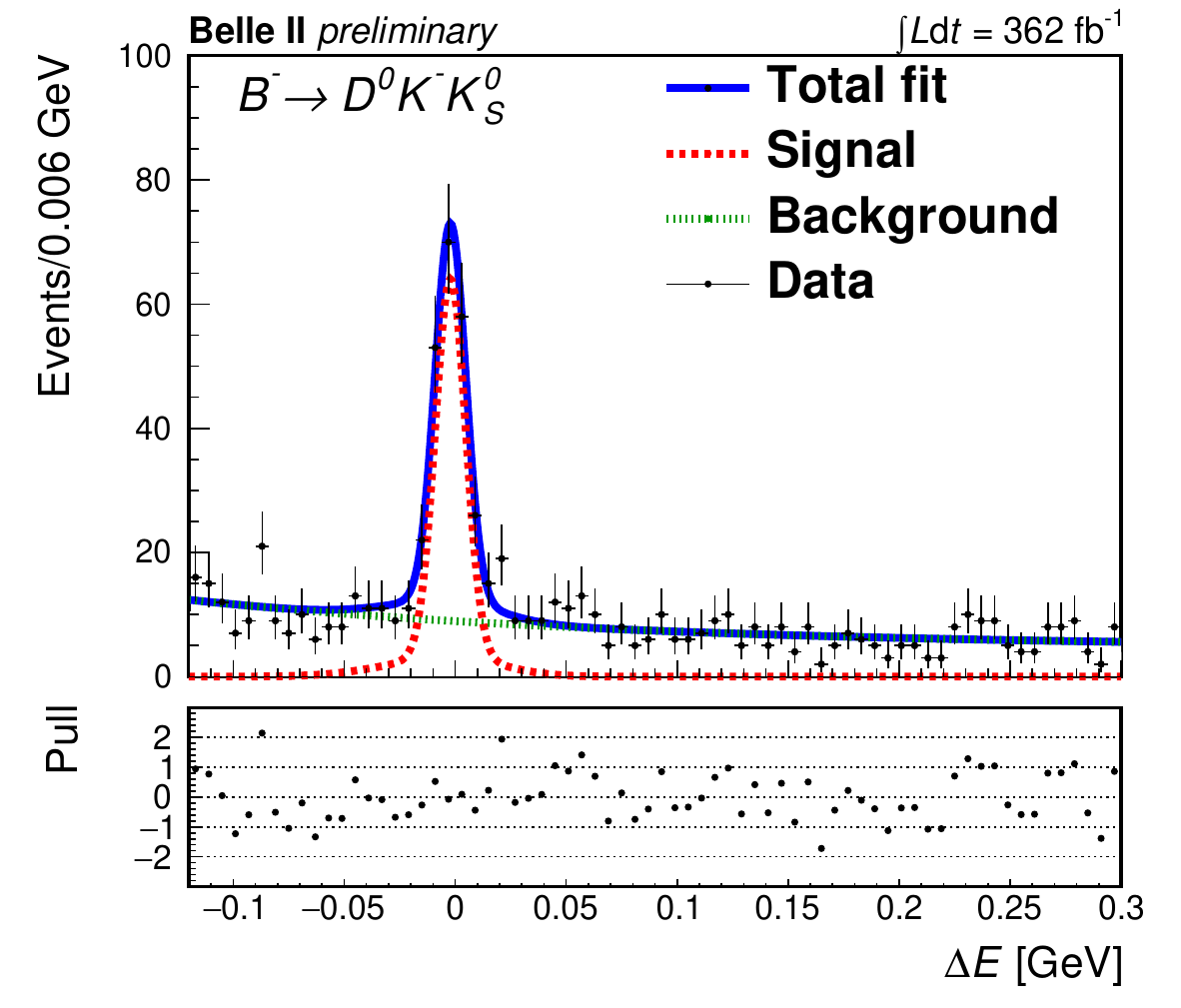}}
\subfigure{\includegraphics[width=0.49\columnwidth]{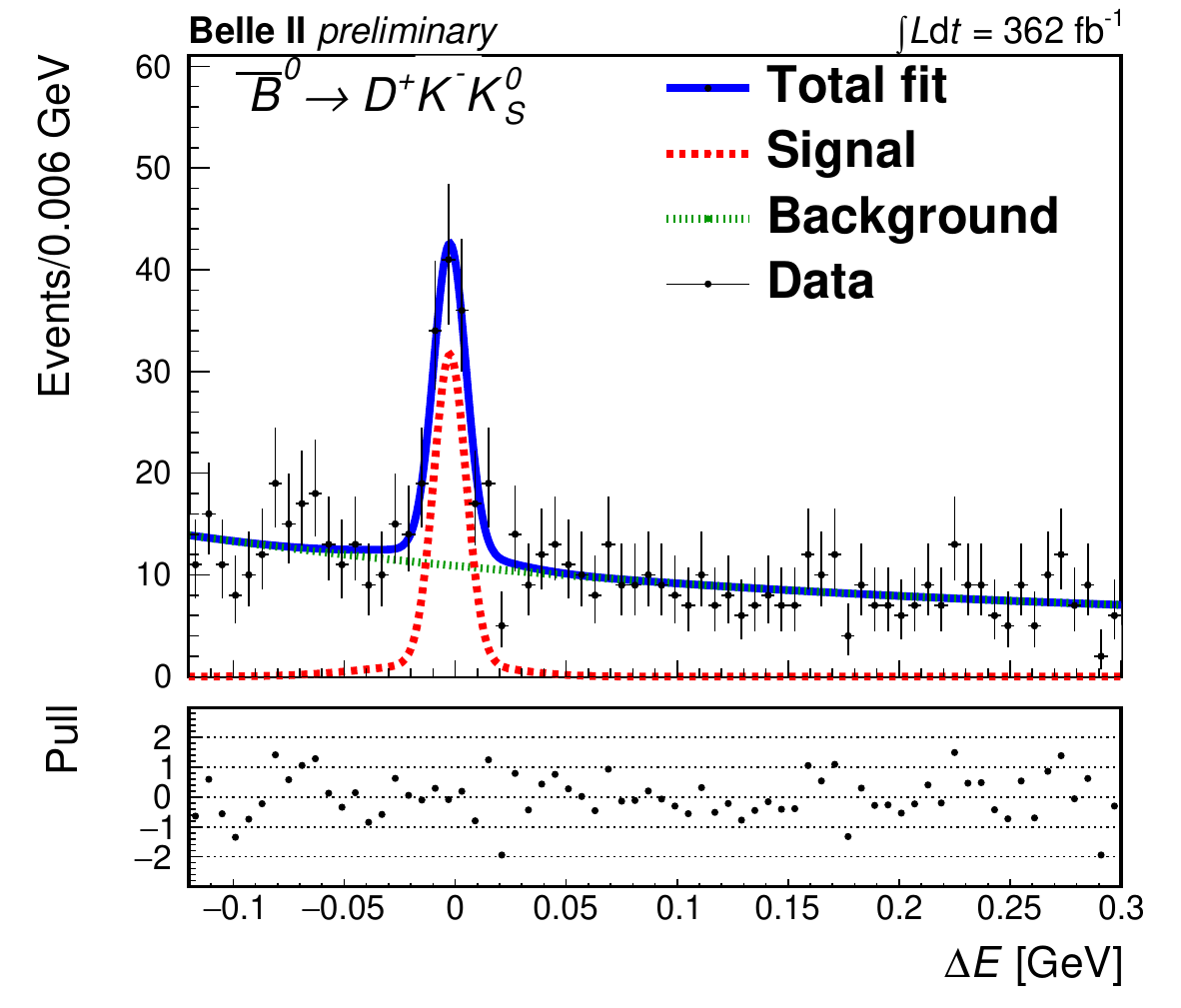}}
\subfigure{\includegraphics[width=0.49\columnwidth]{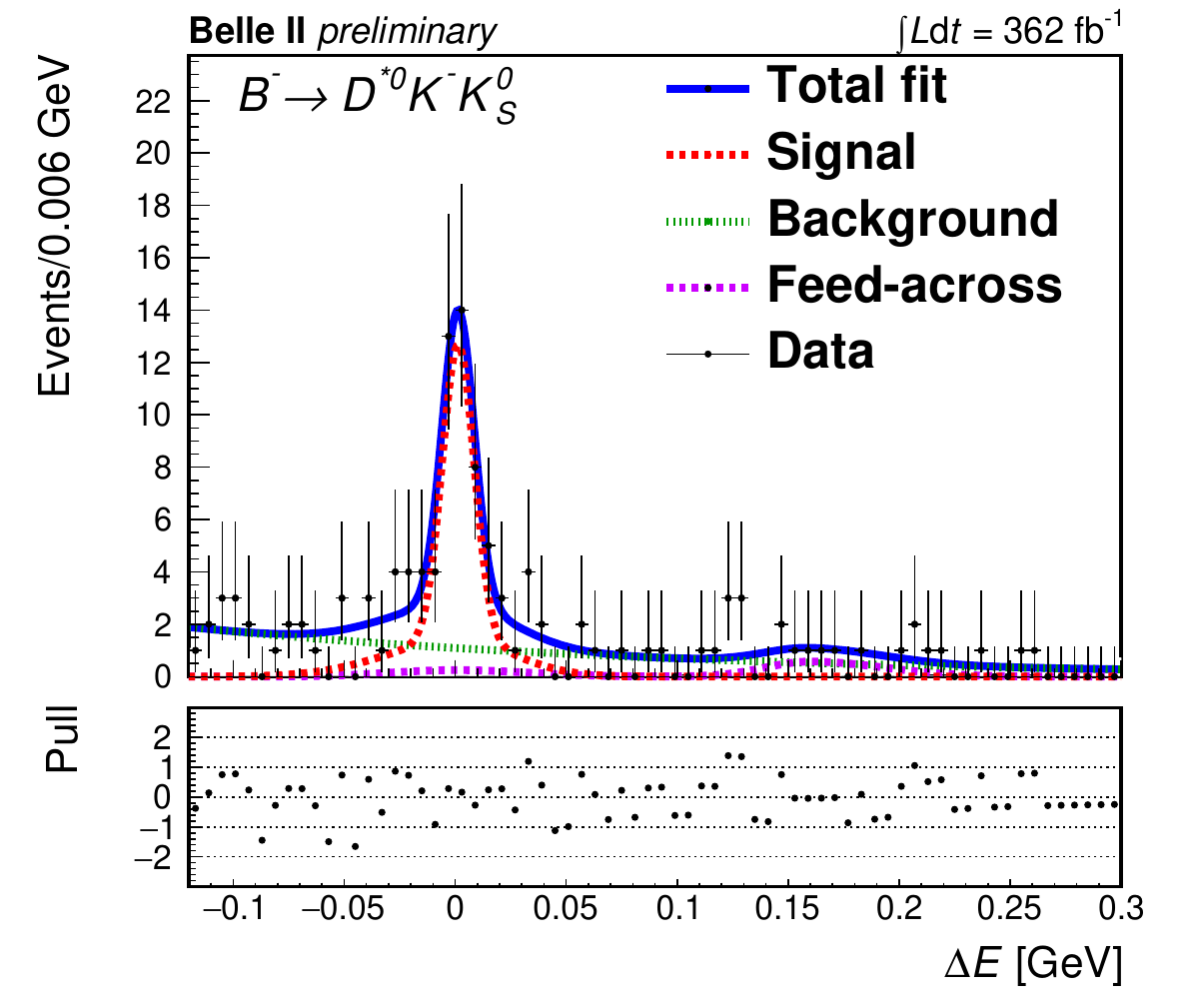}}
\subfigure{\includegraphics[width=0.49\columnwidth]{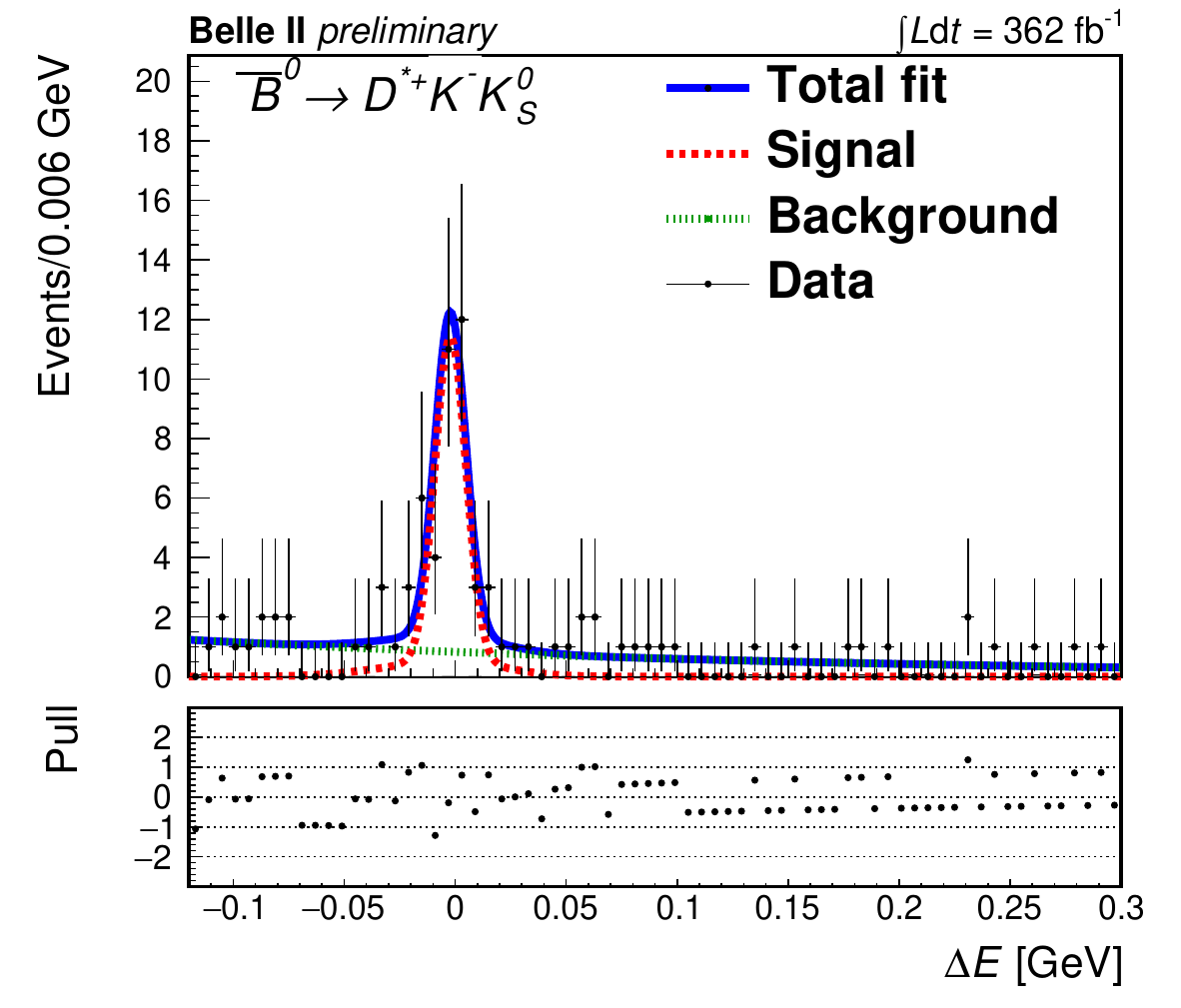}}
\caption{Distribution of $\Delta E$ for $B^-\to D^0K^-K_S^0$ (top left), $\overline B{}^0\to D^+K^-K_S^0$ (top right), $B^-\to D^{*0}K^-K_S^0$ (bottom left), and $\overline B{}^0\to D^{*+}K^-K_S^0$ (bottom right) channels, with the projection of fits overlaid. The fit components are highlighted, and the pulls between the fit and the data are shown below each distribution. \label{fig:deltaE_fit_KS}}
\end{figure}

Belle II reported  a measurement of the branching fractions of four $B^{0, -} \to D^{(*)+,0}K^{-} K_{S}^{0}$~\cite{Belle-II:2023gye} decay modes using $362~\rm fb^{-1}$ of data collected at the $\Upsilon(4S)$ resonance. The event yields are extracted from fits to the distributions of $\Delta E$ to separate signal and background,
and are efficiency-corrected as a function of the invariant mass of the $K^{-}K^{0}_{S}$ system. The $\Delta E$ distributions to Data with fit projections overlaid are shown in Fig.~\ref{fig:deltaE_fit_KS}.
The branching fractions reported to be
\begin{eqnarray}
{\cal B}(B^- \to D^0 K^- K_{S}^0)  &=&  (1.89 \pm 0.16 \pm 0.10) \times 10^{-4},\\
{\cal B}(\bar{B}^0 \to D^+ K^- K_{S}^0)  &=&  (0.85 \pm 0.11 \pm 0.05) \times 10^{-4},\\
{\cal B}(B^- \to D^{*0} K^- K_{S}^0)  &=&  (1.57 \pm 0.27 \pm 0.12) \times 10^{-4},\\
{\cal B}(\bar{B}^0 \to D^{*+} K^- K_{S}^0)  &=&  (0.96 \pm 0.18 \pm 0.06) \times 10^{-4},
\label{eq:belleii_BtoDKK}
\end{eqnarray}
where the first uncertainty is statistical and the second systematic. These results include
the first observation of $\bar{B}^0 \to D^+ K^- K_{S}^0$ $B^- \to D^{*0} K^- K_{S}^0$ $\bar{B}^0 \to D^{*+} K^- K_{S}^0$ decays and a significant improvement in the precision of ${\cal B}(B^- \to D^0 K^- K_{S}^0)$ compared to previous measurements.
Previously, Belle provided the most precise
measurements for $B \to D \pi^+$~\cite{Belle:2021udv, Belle:2021nyg} using its full Belle dataset which corresponds to $772 \times 10^6 B\bar B$ pairs.
\section{Summary}

Search for new sources of $CP$ violation in $B$ meson decays is an interesting but challenging mission. Precise determination of $\phi_3/\gamma$ is very important to quantify the SM sources of $CP$ violation. Precise measurements are already carried out by LHCb and by Belle and Belle II experiments. LHCb by the end of Run 4 with an expected unprecedented amount of data to be collected and Belle II with high integrated luminosity expect to provide a sub-degree precision on $\gamma$. Inputs on charm decays from BESIII with seven times more data plays an important role in achieving this goal.  $B$-meson decays into charmless hadrons offer rich opportunities to test the Standard Model and search for new physics. Rich signatures have been explored using the Run 1 + Run 2 data by LHCb, shedding light into the underlying dynamics responsible for CP violation effects.
The state of the art on the decays of $B$ mesons to purely hadronic final states in QCDF has been discussed. In particular, in the recoil region where the pseudo-two body approach can be applied. The mathematical description demands the introduction of new quantities such as dimeson LCDAs and dimeson form factors which are currently the subject of research using data driven approaches as well as QCD sum rules and lattice QCD. Efforts in the baryonic sector are also ongoing with a broad program ahead.

\section*{Acknowledgements}

We would like to thank the organizers of the CKM 2023 conference for creating a very pleasant and inspiring atmosphere. The work of TH and GTX was supported by the Deutsche Forschungsgemeinschaft (DFG, German Research Foundation) under grant  396021762 - TRR 257 ``Particle Physics Phenomenology after the Higgs Discovery''. Moreover, GTX acknowledges support from the European Union’s Horizon 2020 research and innovation programme under the
Marie Sklodowska-Curie grant agreement No 945422.

\bibliographystyle{amsplain}

\end{document}